\documentclass[aps,pre,twocolumn,superscriptaddress]{revtex4-1}
\usepackage{amsmath,amssymb,graphicx,color,braket,hyphenat,makeidx,xcolor}
\usepackage[ocgcolorlinks,colorlinks=true,linkcolor=blue,citecolor=red,linktocpage=true]{hyperref}

\newcommand{\orange}[1]{{\color{black} #1}}

\begin{document}

\title{Performance of autonomous quantum thermal machines: \\ Hilbert space dimension as a thermodynamical resource}

\author{Ralph Silva}\affiliation{D\'epartement de Physique Th\'eorique, Universit\'e de Gen\`eve, 1211 Gen\`eve, Switzerland}
\author{Gonzalo Manzano}
\affiliation{Departamento de F\'isica At\'omica, Molecular y Nuclear and GISC, Universidad Complutense Madrid, 28040 Madrid, Spain}
\author{Paul Skrzypczyk}\affiliation{H. H. Wills Physics Laboratory, University of Bristol, Tyndall Avenue, Bristol, BS8 1TL, United Kingdom}
\author{Nicolas Brunner}\affiliation{D\'epartement de Physique Th\'eorique, Universit\'e de Gen\`eve, 1211 Gen\`eve, Switzerland}

\begin{abstract}
Multilevel autonomous quantum thermal machines are discussed. In particular, we explore the relation between the size of the machine (captured by Hilbert space dimension), and the performance of the machine. 
Using the concepts of virtual qubits and virtual temperatures, we show that higher dimensional machines can outperform smaller ones. For instance, by considering refrigerators with more levels, lower temperatures can be achieved, as well as higher power. We discuss the optimal design for refrigerators of a given dimension. As a consequence we obtain a statement of the third law in terms of Hilbert space dimension: reaching absolute zero temperature requires infinite dimension. These results demonstrate that Hilbert space dimension should be considered a thermodynamic resource. 
 
\end{abstract}

\maketitle

\section{Introduction}

Autonomous quantum thermal machines function via thermal contact to heat baths at different temperatures, powering different thermodynamic operations without 
any external source of work. For instance, small quantum absorption refrigerators use only two thermal reservoirs, one as a heat source, and the other as a heat sink, in order to cool a system to a temperature lower than that of either of the thermal reservoirs \cite{schulz,palao,linden10,levy,magic}. More generally, autonomous quantum thermal machines represent an ideal platform for exploring quantum thermodynamics \cite{book,review1,review2}, as they allow one to avoid introducing explicitly the concept of work, a notably difficult and controversial issue. The efficiency of these machines has been investigated \cite{schulz,skrzypczyk,correa1,woods}, and quantum effects, such as coherence and entanglement, were shown to enhance their performance \cite{brunner14,correa14, uzdin15,marcus,brask15b,frenzel16}. Also, these machines are of interest from a practical point of view, and several implementations have been proposed \cite{venturelli,chen,mari,bellomo,brask15,mitchison2}. 

More formally, autonomous thermal machines are modelled by considering a set of quantum levels (the machine), some of which are selectively coupled to different thermal baths as well as to an object to be acted upon. Various models of thermal baths and thermal couplings can be considered and formalized via master equations, which usually involves many different parameters, including coupling factors or bath spectral densities, to precisely characterize the machine and its interaction with the environment (see e.g. \cite{marcus}).

Nevertheless, the basic functioning of these machines can be captured in much simpler terms. In particular, the notion of `virtual qubits' and `virtual temperatures' \cite{virtual} (see also \cite{janzing}), essentially associating a temperature to a transition via its population ratio, was developed in order to capture the fundamental limitations of the simplest machines. Therefore, some of the main features of the machine can be deduced from simple considerations about its {\it static} configuration, i.e. without requiring any specific knowledge about the dynamics of the thermalization process induced by contact with the baths.

In the present work we discuss the performance of general thermal machines, involving an arbitrary number of levels. Exploiting the notions of virtual qubits and virtual temperatures, we characterize fundamental limits of such machines, based on its level structure and the way it is coupled to the reservoirs. This allows us to explore the relation between the size of the machine as given by its Hilbert space dimension (or equivalently the number of its available levels), and its performance. We find that machines with more levels can outperform simpler machines. In particular, considering fixed thermodynamic resources (two heat baths at different temperatures), we show that lower temperatures, as well as higher cooling power, can always be engineered using higher dimensional refrigerators. By characterizing the range of virtual qubits and virtual temperatures that can be reached with fixed resources, we propose optimal designs for single-cycle, multi-cycle and concatenated machines featuring an arbitrary number of levels. Furthermore, our considerations lead to a formulation of the third law in terms of Hilbert space dimension of the machine: reaching absolute zero temperature requires infinite dimension. 

The paper is organized as follows. We begin in Sec.~\ref{sec:primitive} by discussing the role of the swap operation as the primitive operation for the functioning of autonomous quantum thermal machines, allowing an extremely simple characterization of their performance in terms of virtual qubits and virtual temperatures. Sec.~\ref{sec:qutrit} is devoted to reviewing the basic functioning of a three-level quantum thermal machine, helping us to identify various {\it resources and} {\it limitations} when optimizing its design. Our general results for higher dimensional thermal machines are presented in Sec.~\ref{sec:results}, where we point out the existence of two different strategies for improving performance. The first strategy consists of adding energy levels to the original thermal cycle, and is analyzed in detail in Sec.~\ref{sec:single-cycle}, while the extension to the case of multi-cycle machines in presented in Sec.~\ref{sec:multi-cycle}. The second strategy, based upon concatenating qutrit machines, is analyzed in Sec.~\ref{sec:concatenated}. Furthermore, in Sec.~\ref{sec:thirdlaw} we discuss the third law of thermodynamics in terms of Hilbert space dimension, while Sec.~\ref{sec:dynamics} is devoted to characterizing the trade-off between the power and speed of operation of the thermal machine, given an explicit model of thermalization. Finally, our conclusions are presented in Sec.~\ref{sec:conclusions}.

\section{The primitive operation} \label{sec:primitive}

Generally speaking, the working of an autonomous quantum thermal machine can be divided into two steps which are continuously repeated. For clarity, we discuss the case of a fridge powered by two thermal baths at different temperatures. In the first step, a temperature colder than the cold bath is engineered on a subspace of the machine, i.e. on a subset of the levels comprising the machine. This can be done by selectively coupling levels in the machine to the thermal baths. The second step consists in interacting the engineered subspace with an external physical system to be cooled. We will consider a pair of levels of the machine to constitute our engineering subspace, the population ratio of which can be tuned in order to correspond to a cold temperature. Here we shall refer to this pairs of levels as the `virtual qubit', and its associated 
temperature as its `virtual temperature' \cite{virtual}. Typically the virtual qubit is chosen to be resonant with the system to be cooled in order to avoid non energy conserving interactions. Notably, the swap operation between the virtual qubit and the external physical system, can thus be considered as the primitive operation of quantum fridges, and more generally of all quantum thermal machines.

Let us consider a machine comprised of $n$ levels, with associated Hilbert space $\mathcal{H}$ such that ${\rm dim}\mathcal{H} = n$, 
and Hamiltonian $H_{\rm M}$. Within this machine, we will refer to any pair of levels ($\ket{k}$ and $\ket{l}$) as a {\it transition}, denoted 
$\Gamma_{k,l}$. Among the $n(n-1)/2$ possible transitions, we focus our attention on a particular pair of levels $\ket{i}$ and $\ket{j}$ with 
populations $\lambda_i$ and $\lambda_j$ and energies $E_i$ and $E_j > E_i$. Assume the transition $\Gamma_{i, j}$ is coupled to the external 
system to be cooled, hence represening the virtual qubit. Here it will be useful to introduce two quantities to fully 
characterize the virtual qubit, namely its normalization $N_{\rm v}$ and its (normalized) bias $Z_{\rm v}$ defined by
\begin{align}
N_{\rm v} ~:=~ \lambda_i + \lambda_j \quad  \quad
Z_{\rm v}  ~:=~ \frac{\lambda_i - \lambda_j}{N_{\rm v}}.
\end{align} 
As we focus here on the case where the density operator of the machine is diagonal in the energy basis \cite{footnote1}, we may define 
its temperature, i.e. the virtual temperature, via the Gibbs relation $\lambda_j = \lambda_i e^{- E_{\rm v}/ k_\mathrm{B} T_{\rm v}}$. That is
\begin{align}
T_{\rm v} := \frac{E_{\rm v}}{k_\mathrm{B}} \ln\frac{\lambda_j}{\lambda_i} 
\end{align}
where we defined $E_{\rm v} ~:=~ E_j -E_i$ as the energy gap of the virtual qubit. The virtual temperature is then monotonically related 
to the above introduced bias by
\begin{equation}
Z_{\rm v} = \tanh(\beta_{\rm v} E_{\rm v}/2)
\end{equation}
where $\beta_{\rm v} = 1/k_\mathrm{B} T_{\rm v}$ is the inverse virtual temperature. Notice that $-1 \leq Z_{\rm v} \leq 1$, where the lower bound represents a virtual 
qubit with complete population inversion ($\beta_{\rm v} \rightarrow - \infty$) and the upper bound correspond to the virtual qubit in its ground state 
$\ket{i}$ ($\beta_{\rm v} \rightarrow 0$).

Next, we interact the virtual qubit with the physical system via the swap operation. For simplicity, the physical system is taken here to be a qubit 
with energy gap $E_{\rm v}$, hence resonant with the virtual qubit. We denote the levels of the physical system by $\ket{0}$ and $\ket{1}$, with corresponding populations $p_0$ and $p_1$, and hence bias  $Z_{\rm s}= p_0 - p_1$ (note that $N_{\rm s}=1$). The swap operation is given by
\begin{align}
U = \mathbb{I} &- \ket{i,1}\bra{i,1} - \ket{j,0}\bra{j,0}~ + \nonumber \\ 
&+ \ket{i,1}\bra{j,0} + \ket{j,0}\bra{i,1}.
\end{align}
The effect of the swap operation is to modify the bias of the physical system, which changes from $Z_{\rm s}$ to 
\begin{equation}\label{swapeffect}
	Z_{\rm s}^\prime = N_{\rm v} Z_{\rm v} + (1-N_{\rm v})Z_{\rm s}.
\end{equation}
The above equation can be intuitively understood as follows. With probability $N_{\rm v}$, the virtual qubit is available (i.e. the machine is 
in the subspace of the virtual qubit), and the swap replaces the initial bias of the system with the bias of the virtual qubit. With the 
complementary probability, $1-N_{\rm v}$, the virtual qubit is not available, hence the swap cannot take place and the bias of the system remains unchanged. Consequently, the virtual temperature fundamentally limits the temperature the external system can reach. A complete derivation of Eq.~\eqref{swapeffect} can be found in Appendix \ref{AppSwap}.

Finally, it is worth noticing that the virtual qubit must be refreshed in order to ensure the continuous operation of the machine. Indeed, after interaction with the system, the virtual qubit is left with the initial bias of the system, $Z_{\rm s}$, and must be therefore reset to the desired bias, $Z_{\rm v}$, in order to continue operating.

Given the above perspective on the working of quantum thermal machines, two different directions to improve the performance of a machine \orange{emerge}. 
The first consists in optimizing the properties of the virtual qubit ($N_{\rm v} $ and $Z_{\rm v} $) \orange{in order to achieve the desired bias $Z_{\rm s}^\prime$ in 
the external system ($Z_{\rm s}^\prime \rightarrow 1$ in the case of a fridge)}, which represent the {\it statics} of the 
machine. The second consists in optimizing the dynamics of the machine, in particular the rate of interaction with the \orange{external} system and the 
rate at which the virtual qubit is refreshed by contact with the thermal baths. Crucially, whereas the dynamics is model dependent, the 
statics are model independent, and hence universal properties of the machine. 

In the following sections, we shall see how the performance of thermal machines can be optimized in the presence of natural constraints, \orange{such as limits on the available energy gaps or 
on the dimension of its Hilbert space}. Our focus will primarily be on the statics: we will see that increasing the number of levels of the machine 
will allow for increased performance (for instance to be able to cool to lower temperatures). However, in the last sections, we will move beyond purely 
static considerations, and discuss the interplay between statics and dynamics. Again we find that machines with more levels can lead to 
enhanced performance.

\section{Warm-up: qutrit machine} \label{sec:qutrit}

\begin{figure}[b!]
\includegraphics[width=0.8\linewidth]{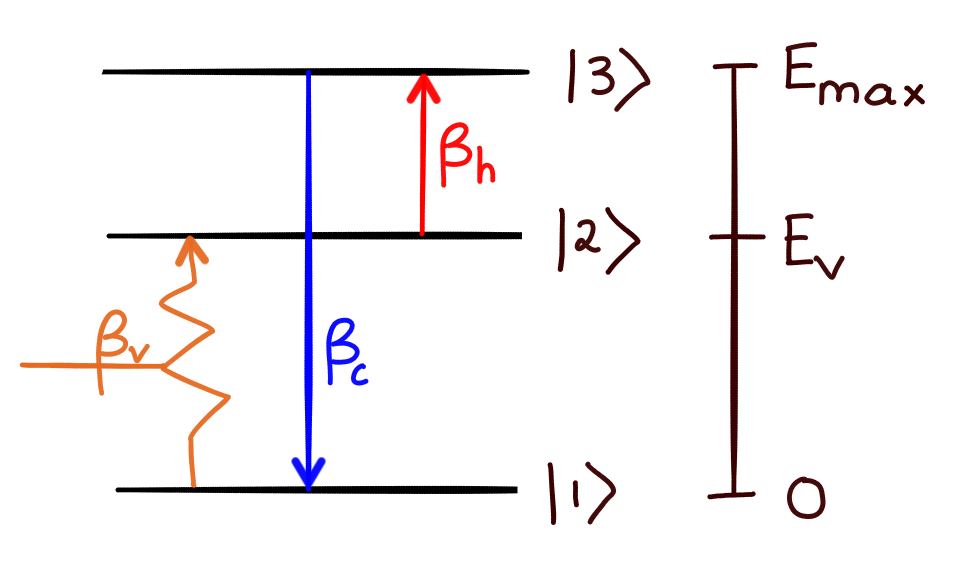}
\caption{The smallest possible fridge comprising three energy levels. Throughout this paper, couplings to $\beta_{\rm c}$ will be denoted by (blue) downward arrows, couplings to $\beta_{\rm h}$ by (red) upward arrows, and the virtual qubit by an (orange) arrow in the direction consistent with the machine (upward for the fridge, downward for the engine). \label{qutrit}}
\end{figure}

In order to better ilustrate the main concepts, we start our analysis with the smallest possible quantum thermal machine, 
comprising only three energy levels 
$\ket{1}$, $\ket{2}$ and $\ket{3}$, working between two thermal baths at different temperatures. This machine can be operated as a fridge 
or as a heat engine depending on which transitions are coupled to the hot and cold baths. For simplicity, our presentation will focus on the 
former (see Fig.~\ref{qutrit}). In this case, the transition $\Gamma_{1,3}$ is coupled to the cold bath at inverse temperature $\beta_{\rm c}$, while 
transition $\Gamma_{2,3}$ is coupled to the hot bath at $\beta_{\rm h} < \beta_{\rm c}$. Finally, the transition $\Gamma_{1,2}$ is choosen to be 
the virtual qubit.

The operation of the qutrit fridge can be understood as a simple thermal cycle:
\begin{equation}
	\ket{2} \xrightarrow{\beta_{\rm h}} \ket{3} \xrightarrow{\beta_{\rm c}} \ket{1}.
\end{equation}
in which a quantum of energy $\Delta E_{23} \equiv E_3 - E_2$ is adsorbed from the hot bath making the machine jump from state $\ket{2}$ to $\ket{3}$, 
followed by a jump from $\ket{3}$ to $\ket{1}$ while emiting a quantum of energy $\Delta E_{13}$ to the cold bath. The cycle is closed 
by swap of the virtual qubit, $\Gamma_{1,2}$, with the external qubit to be cooled as described in Sec.~\ref{sec:primitive}. This cycle 
involves 3 states, and is thus of length 3. It represents the basic building block of the machine.

The fact that transitions $\Gamma_{1,3}$ and $\Gamma_{2,3}$ are coupled to baths at different temperatures will allow us to control the (inverse) 
temperature of the virtual qubit, $\beta_{\rm v}$. While there exist many different possible models for representing the coupling to a thermal bath, 
the only feature that we will consider here is that, after sufficient time, each transition connected to a bath will thermalize. 
That is, in the steady-state of the machine, the population ratio of a transition $\Gamma_{i,j}$ coupled to a thermal bath, will be equal to 
$e^{- \Delta E_{ij} \beta_{\rm bath}}$, where $\Delta E_{ij}$ is the energy gap of the transition, and $\beta_{\rm bath}$ 
the inverse temperature of the bath.
Under such conditions, the inverse temperature of the virtual qubit and its norm are given by
\begin{align}\label{betavfridge0}
	\beta_{\rm v} &= \beta_{\rm c} + (\beta_{\rm c} - \beta_{\rm h}) \left( \frac{ \Delta E_{13}}{E_{\rm v}} -1 \right), \\ \label{normfridge0}
	 N_{\rm v} &= \frac{1 + e^{-\beta_{\rm v} E_{\rm v}}}{1 + e^{-\beta_{\rm v} E_{\rm v}} + e^{-\beta_{\rm c} \Delta E_{13}}} 
\end{align}
where $E_{\rm v} \equiv \Delta E_{12}$ is the virtual qubit energy gap, chosen to match the energy gap of the qubit to be cooled. Note that we have 
$\beta_{\rm v}  > \beta_{\rm c}$ (since $\Delta E_{13} > E_\mathrm{v}$), implying that the machine works as a refrigerator.

At this point, one can already identify various {\it resources} for the control of the virtual temperature $\beta_{\rm v}$. The first is the range of 
available temperatures, captured by $\beta_{\rm c}$ and $\beta_{\rm h}$. The second is the largest energy gap, $\Delta E_{13}$ coupled to a thermal bath. Clearly if 
$\Delta E_{13}$ is unbounded, then we can cool arbitrarily close to absolute zero, i.e. $\beta_{\rm v}\rightarrow \infty$ as $\Delta E_{13} \rightarrow \infty$ 
while $N_{\rm v} \rightarrow 1$, implying $Z_{\rm s}^\prime \rightarrow 1$, c.f. Eq.~\eqref{swapeffect}. However, it is reasonable to impose a bound on this quantity, which we label $E_{\rm max}$. From physical considerations, one expects that thermal effects play role only up to a certain energy scale. In general, a thermal bath is characterized by a spectral density with a cutoff for high frequencies. This implies the existence of an energy above which there exist a negligible number of systems in the bath. 
In any case, the coldest achievable temperature given this maximum energy is then given by
\begin{align}\label{betavfridge1}
	\beta_{\rm v} &= \beta_{\rm c} + (\beta_{\rm c} - \beta_{\rm h}) \left( \frac{ E_{\rm max}}{E_{\rm v}} -1 \right).
\end{align}

As mentioned above, the qutrit machine can also work as a heat pump or heat engine, if one switches the hot and cold baths. Imposing again a maximum 
energy gap coupled to a bath we obtain the following lower bound in the inverse virtual temperature
\begin{align}\label{betavengine1}
	\beta_{\rm v} &= \beta_{\rm h} - (\beta_{\rm c}-\beta_{\rm h}) \left( \frac{  E_{\rm max}}{E_{\rm v}} -1 \right) .
\end{align}
Notice that in this case $\beta_{\rm v} < \beta_{\rm h}$. Moreover, when $\beta_\mathrm{c}/(\beta_\mathrm{c} - \beta_\mathrm{h}) < E_\mathrm{max}/E_\mathrm{v}$, then $\beta_\mathrm{v} < 0$, and the machine transitions from a heat pump to a heat engine.

\section{Summary of results} \label{sec:results}

We have seen that imposing a bound on the maximum energy gap the performance of the simplest qutrit machine becomes limited through 
the range of accessible virtual temperatures. The general question investigated below is whether these limits 
can be overcome. That is, can we engineer colder temperatures (or hotter ones, as well as achieving population inversion) by using 
more sophisticated machines? 

\begin{figure}[t!]
\includegraphics[width=\linewidth]{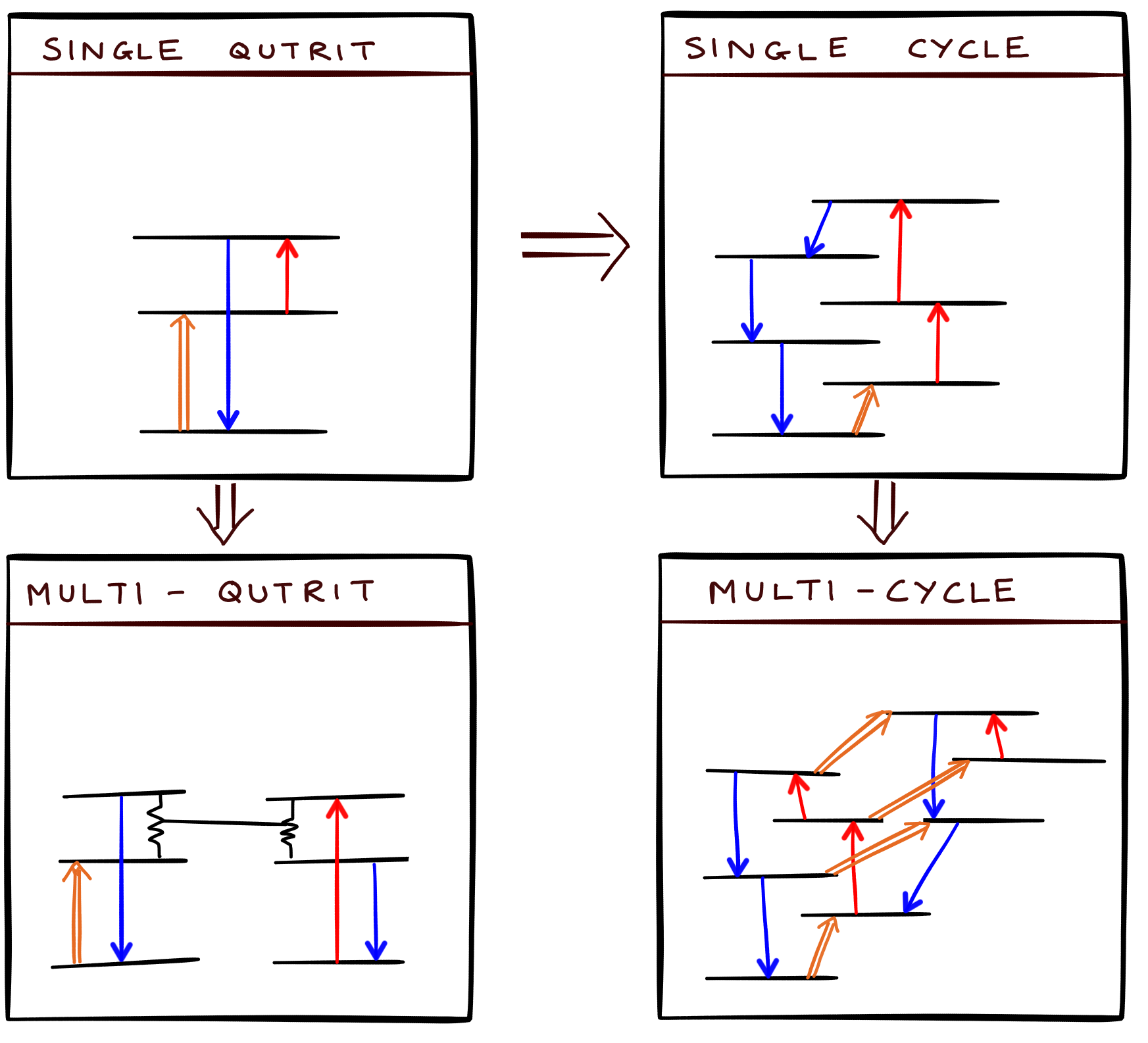}
\caption{Sketch of machines discussed in the present work. We consider several generalizations of the simplest qutrit machine (top left). We first discuss single cycle machine (top right), which can then be extended to multi-cycle machines (bottom right). Second, we study concatenated qutrit machines (bottom left). \label{outlook}}
\end{figure}

Clearly, in order to optimize the effect the machine has on the physical system, there are two important features the virtual qubit should have 
following Eq.~\eqref{swapeffect}. First, it should have a high bias $Z_{\rm v}$. Second, the norm $N_\mathrm{v}$ should be as close to one as possible. 
Below we discuss different classes of multilevel machines, and investigate the range of available virtual qubits as a function 
of the number of levels $n$ of the machine. First we will see that the range of accessible virtual temperatures (or equivalently
bias $Z_{\rm v}$) increases as $n$ increases. Hence machines with more levels allow one to reach lower temperatures, given fixed thermal resources. 
However, this usually comes at the price of having a relatively low norm $N_{\rm v}$ for the virtual qubit, which is clearly a detrimental feature. 
Nevertheless we will see that it is always possible to bring the norm back to one by adding extra levels.

\begin{figure}[t!]
\includegraphics[width=\linewidth]{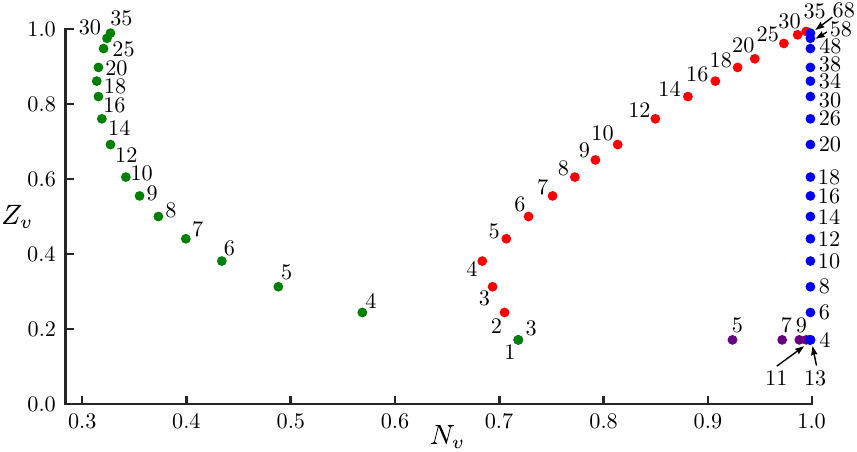}
\caption{Performance of machines as a function of dimension. The accessible virtual qubit, characterized by the bias $Z_{\rm v}$ and the norm $N_{\rm v}$, is 
shown for single cycle machine (green dots), multi-cycle machine (blue dots), and concatenated qutrit machines (red dots). As a comparison we 
also show the machines discussed in Ref. \cite{CorreaNv} (purple dots). The dimension of the machine (i.e. the number of levels) is indicated next to each point, for all machines except the qutrit; there, the number $k$ of concatenated machines is given (hence the dimension is exponentially larger, $3^k$).  \label{mainres}}
\end{figure}

We discuss two natural ways to generalize the qutrit machines to more levels, sketched in Fig.~\ref{outlook}. The first one 
consists in adding levels and thermal couplings in order to extend the length of the thermal cycle. In other words, while the qutrit machine 
represents a machine with one cycle of length three, we now consider machines with a single cycle of length $n$. This will allow us to 
improve both the bias and the normalization of the virtual qubit. We first characterize the optimal single-cycle machine, 
which in the limit of large $n$, approaches perfect bias (i.e. zero virtual temperature, or perfect 
population inversion). However, while the norm $N_{\rm v}$ does not vanish, it is bounded away from one in this case. We then show how the 
norm can be further increased to one by extending the optimal single-cycle machine to a multi-cycle machine. This procedures requires 
the addition of $n-2$ levels, while maintaining the same bias. 
In Fig.~\ref{mainres} we show the range of available virtual qubits (as characterized by its norm $N_{\rm v}$ and bias $Z_{\rm v}$) as a function of 
the number of levels $n$, for single cycle machines (green dots) and multi-cycle machines (blue dots).

Next, we follow a second possibility which consists in concatenating $k$ qutrit machines. The main idea is that the hot bath is now 
effectively replaced by an even hotter bath/source of work, engineered via the use of an additional qutrit heat pump/engine. In the limit of $k$ large, we can also 
approach perfect bias and the norm tends to one (see red dots on Fig.~\ref{mainres}), similarly to the multi-cycle machine. It is 
however worth mentioning that in this case the machine has now $n=3^k$ levels, while the multi-cycle machine used only a number of 
levels linear in $n$. 

The above results, which are summarized in Fig.~\ref{mainres}, clearly demonstrate that machines with a larger Hilbert space can outperform 
smaller ones, which implies that the Hilbert space dimension should be considered a thermodynamical resource. Note that, for clarity, 
results are generally discussed for the case of fridges, but hold also for heat engines {\it mutatis mutandis}. 

\section{Single-cycle machines}\label{sec:single-cycle}

We start by discussing thermal machines featuring an arbitrary number of levels, $n$, but only a single thermal cycle. We define a 
{\it $n-$level (thermal) cycle machine} as a quantum system with Hilbert space $\mathcal{H}$ of dimension $n$, and 
Hamiltonian $H= \sum_{j=1}^n E_j \ket{j}\bra{j}$, where every transition $\Gamma_{j,{j+1}}$, is coupled to a thermal bath. It is worth mentioning that the levels $\{ \ket{j}\}$, with $1\leq j\leq n$, are not necessarily ordered with respect to its associated 
energies $E_j$. We further denote the energy gap of the transition $\Gamma_{j,{j+1}}$ as $\Delta E_{j,{j+1}} = E_{{j+1}} - E_{j}$, 
and the temperature of the bath coupled to this transition is labelled as $\beta_{j,{j+1}}$. 
We choose the transition $\Gamma_{1,n}$ to correspond to the virtual qubit of the machine, whose energy gap, $E_{\rm v}$, 
obeys the following consistency relation
\begin{equation}\label{virtualgap}
	E_{\rm v} = \sum_{j=1}^{n-1} E_{{j+1}} - E_{j} = \sum_{j=1}^{n-1} \Delta E_{j,{j+1}}.
\end{equation}

In the absence of any additional couplings, the machine approaches a steady state, as each transition tends to equilibrate 
with the thermal bath to which it is coupled. We notice that each level is involved in at least one thermal coupling. 
This implies that the density matrix of the steady state must be diagonal in the energy basis, as all off-diagonal elements decay 
away due to the thermal interactions. Additionally, the populations of the two levels in each transition are given by the Gibbs ratio 
corresponding to the temperature of the bath. Labeling the population of the $\ket{j}$ state as $p_{j}$, we have
\begin{equation}
	\frac{p_{{j+1}}}{p_{j}} = e^{-\beta_{j,{j+1}} \Delta E_{j,{j+1}}} \quad \text{for  } 1 \leq j \leq n-1.
\end{equation}
The above $n-1$ thermal couplings determine the ratios between all of the populations $\{p_{j}\}$. Together with the normalization 
condition $\sum_j p_{j}=1$, this completely determines the steady state of the machine \cite{footnote3}.
The virtual temperature corresponding to transition $\Gamma_{1, n}$ can hence be obtained from
\begin{align}
	e^{-\beta_{\rm v} E_{\rm v}} &= \frac{p_{n}}{p_{1}} = \frac{p_{n}}{p_{n-1}} \frac{p_{n-1}}{p_{n-2}} ~...~ \frac{p_{2}}{p_{1}}, 
\end{align}
leading to
\begin{align}
	\beta_{\rm v} &= \sum_{j=1}^{n-1} \beta_{j,{j+1}} \frac{\Delta E_{j,{j+1}}}{E_{\rm v}}. \label{cyclebetav}
\end{align}
Similarly one may calculate the norm of the virtual qubit,
\begin{align}\label{cycleNv}
	N_{\rm v} &= \left( \frac{1 + e^{-\beta_{\rm v} E_{\rm v}}}{1 + \sum_{j=1}^{n-1} \prod_{k=1}^{k=j} e^{-\beta_{k, {k+1}} \Delta E_{k, {k+1}}}} \right).
\end{align}

We are interested in the best single cycle machine, that is, the one which using a limited set of {\it resources}, achieves the largest change in bias of the system acted upon, $Z_{\rm s}' -Z_{\rm s}$, as given in Eq.~\eqref{swapeffect}). This corresponds to the one that achieves the largest possible 
bias, $Z_{\rm v}$, together with the largest norm, $N_{\rm v}$, given this optimized bias. In what follows we determine the optimal single cycle machine with $n$ levels, given bath temperatures and bound on the energy of a coupled transition $E_{\rm max}$.

\subsection{Optimal single-cycle machine}

\begin{figure}[b!]
\includegraphics[width=0.9\linewidth]{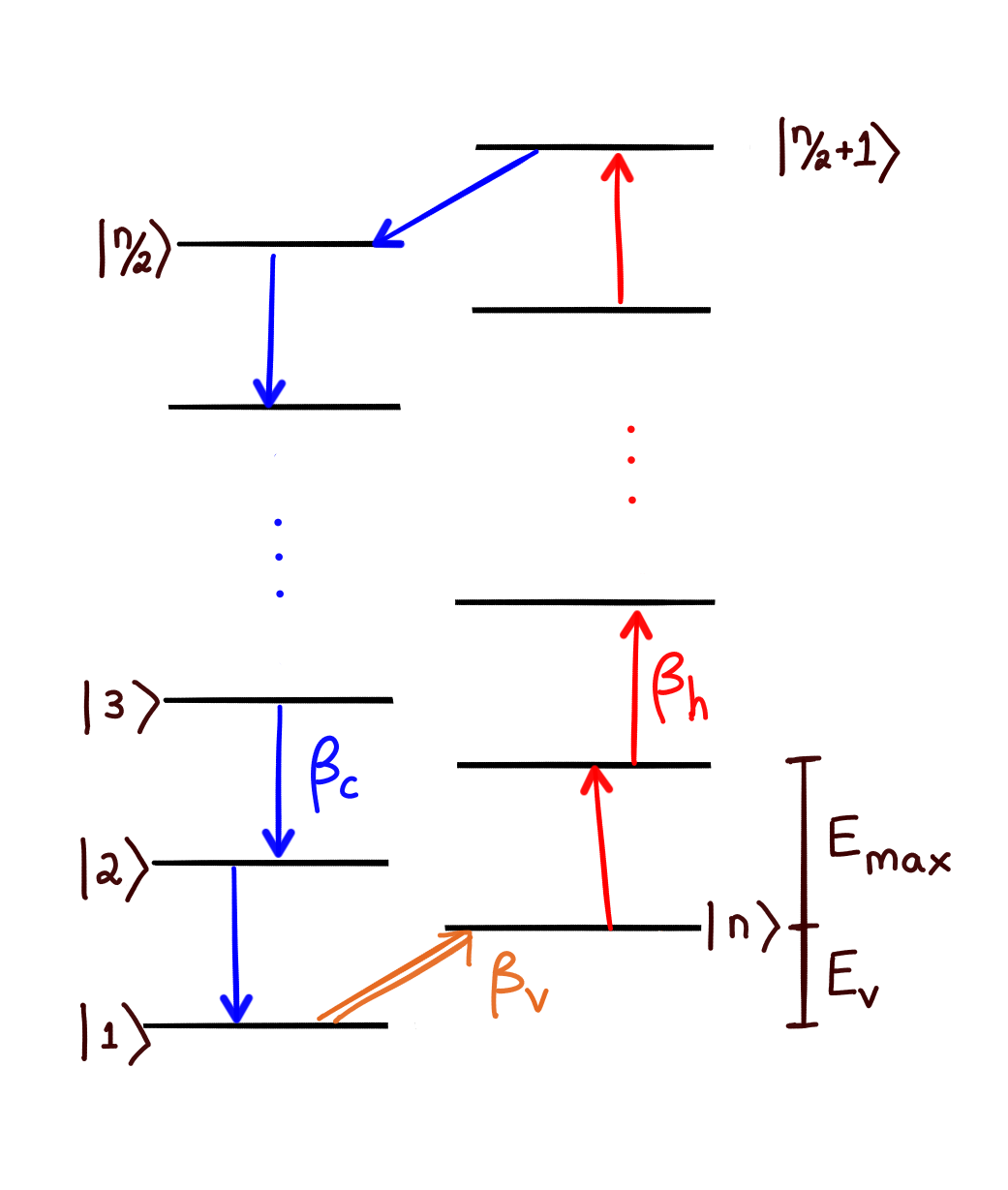}
\caption{Sketch of the optimal single-cycle refrigerator, for an even number of levels $n$.} \label{optimalcycle}
\end{figure}

The optimal arbritrary single cycle fridge, sketched in Fig.~\ref{optimalcycle}, has a rather simple structure. All but one of its transitions are at 
the maximal allowed energy, $E_{\rm max}$. Roughly, the first half of the transitions (starting from the upper state of the virtual qubit) are all 
connected to the hot bath, while the second half of the transitions are connected to the cold bath. A complete proof of optimality can be 
found in Appendix \ref{AppOptTech}. Furthermore, explicit expressions for the inverse virtual temperature and norms in this case can be easily obtained from Eqs.(\ref{cyclebetav}) and (\ref{cycleNv}). 
For the case of refrigerator with an even number of levels $n$, they read 
\begin{eqnarray}\label{betav}
\beta_{\rm v}^{(n)} &=&  \beta_{\rm c} + \left( \beta_{\rm c} - \beta_{\rm h} \right) \left( \frac{n}{2} -1 \right) \frac{E_{\rm max}}{E_{\rm v}} \\ \label{Nv}
N_{\rm v}^{(n)} &=&  \frac{ 1 + e^{-\beta_{\rm v}^{(n)} E_{\rm v}} }{Ê   \frac{  1 - e^{-\frac{n}{2} \beta_{\rm c} E_{\rm max}} }{1 - e^{-\beta_{\rm c} E_{\rm max}}  } + e^{-\beta_{\rm v}^{(n)} E_{\rm v}}  \frac{ 1 - e^{-\frac{n}{2} \beta_{\rm h} E_{\rm max}}  }{ 1 - e^{-\beta_{\rm h} E_{\rm max}}  } },
\end{eqnarray}
while the complete results for all $n$, and heat engines are given, respectively, in Appendices \ref{AppOptTech} and \ref{switching}.

Let us now discuss the performance of the optimal machine. As becomes apparent from Eq.~\eqref{betav}, the number of levels $n$ 
is clearly a thermodynamical resource, as it allows to reach colder temperatures. Indeed, one finds that the virtual temperature is 
improved by a fixed amount whenever two extra levels are added,
\begin{equation}
	\left( \beta_{\rm v}^{(n+2)} - \beta_{\rm v}^{(n)} \right) E_{\rm v} =  \left( \beta_{\rm c} - \beta_{\rm h} \right) E_{\rm max}.
\end{equation}
This relation encapsulates the interplay between the resources involved in constructing a quantum thermal machine - the range of available thermal 
baths $\{\beta_{\rm c}, \beta_{\rm h}\}$, the range of thermal interactions ($E_{\rm max}$), and the number of levels $n$. Remarkably, as the 
inverse virtual temperature $\beta_{\rm v}$ increases linearly with $n$, one can engineer a virtual temperature arbitrarily close to absolute zero. 
Similarly, for a heat engine, one can obtain a virtual qubit with arbitrarily close to perfect population inversion. This is possible because as $n$ increases, the norm of the virtual qubit does not decrease arbitrarily, but remains bounded below away from zero. Indeed from Eq.~\eqref{Nv}, the norm asymptotically approaches a finite value
\begin{equation}
	\lim_{n\rightarrow\infty} N_{\rm v}^{(n)} =  \left( 1 - e^{-\beta_{\rm c} E_{\rm max}} \right),
\end{equation}
which is, interestingly, independent of both $\beta_{\rm h}$ and $E_{\rm v}$.

Finally, we briefly comment on the efficiency (also often referred to as the \emph{coefficient of performance} (COP)) of the optimal single cycle machine. Here we adopt the standard definition of the efficiency 
of an absorption refrigerator, that is, the ratio between the heat extracted from the object to be cooled and the heat extracted from the 
hot bath. This can be easily calculated by looking at a single complete cycle of the machine. Imagine that a quantum $E_{\rm v}$ 
of heat is extracted from the external qubit, in the jump $\ket{1} \rightarrow \ket{n}$ produced by the swap operation. To complete the cycle, the following sequence of jumps must necessarily occur:
\begin{equation}
 \ket{n} \xrightarrow{\beta_{\rm h}} ... \xrightarrow{\beta_{\rm h}} \ket{n/2 +1} \xrightarrow{\beta_{\rm c}} \ket{n/2} \xrightarrow{\beta_{\rm c}} ...  \xrightarrow{\beta_{\rm c}} \ket{1}
\end{equation}
where $n/2 - 1$ energy quanta $E_{\rm max}$ of heat are adsorbed from the hot bath while releasing $n/2 - 1$ quanta $E_{\rm max}$ and one quantum $E_{\rm v}$
of heat to the cold bath. The efficiency is hence given by:
\begin{align}\label{efficiency}
	\eta_\mathrm{fridge}^{(n)} = \frac{E_{\rm v}}{\left( \frac{n}{2} -1 \right) E_{\rm max}}  = \frac{\beta_{\rm c} - \beta_{\rm h}}{\beta_{\rm v}^{(n)} - \beta_{\rm c}}.
\end{align}
where the second equality follows by exploiting Eq.~\eqref{betav} (see Appendix D). Crucially, Eq.~\eqref{efficiency} corresponds to Carnot efficiency for an endoreversible absorption refrigerator that is extracting heat from a bath at the temperature $\beta_{\rm v}^{(n)} \geq \beta_{\rm c} \geq \beta_{\rm h}$. That is, if the object to be cooled (now an external bath) is infinitesimally above the temperature of the virtual qubit (such that the virtual qubit cools it down by an infinitesimal amount), then the efficiency (COP) of this process approaches the Carnot limit.

Note that such absorption refrigerators have the property that the COP drops as the temperature of the cold reservoir drops. In the present case, since $\beta_\mathrm{v}^{(n)}$ drops linearly with $n$, so too does the efficiency of the machine. Intuitively, this makes sense,  
since the amount of heat drawn from the hot bath (per cycle) increases linearly with $n$, while the heat extracted from the cold bath 
remains constant (see Fig.~\ref{optimalcycle}). 

\section{Multi-cycle machines}\label{sec:multi-cycle}

We have seen that the optimal single cycle machine can enhance the virtual temperature by increasing the number of levels $n$. Basically, 
this comes at the price of having the norm $N_{\rm v}$ relatively low, which is clearly a detrimental feature. Hence, it is natural to ask if, 
by adding levels, the norm can be brought back to unity while keeping the same virtual temperature. Below we will see that this is always possible, and in fact, requires only (roughly) twice the number of levels. 

\begin{figure}[b]
\includegraphics[width=\linewidth]{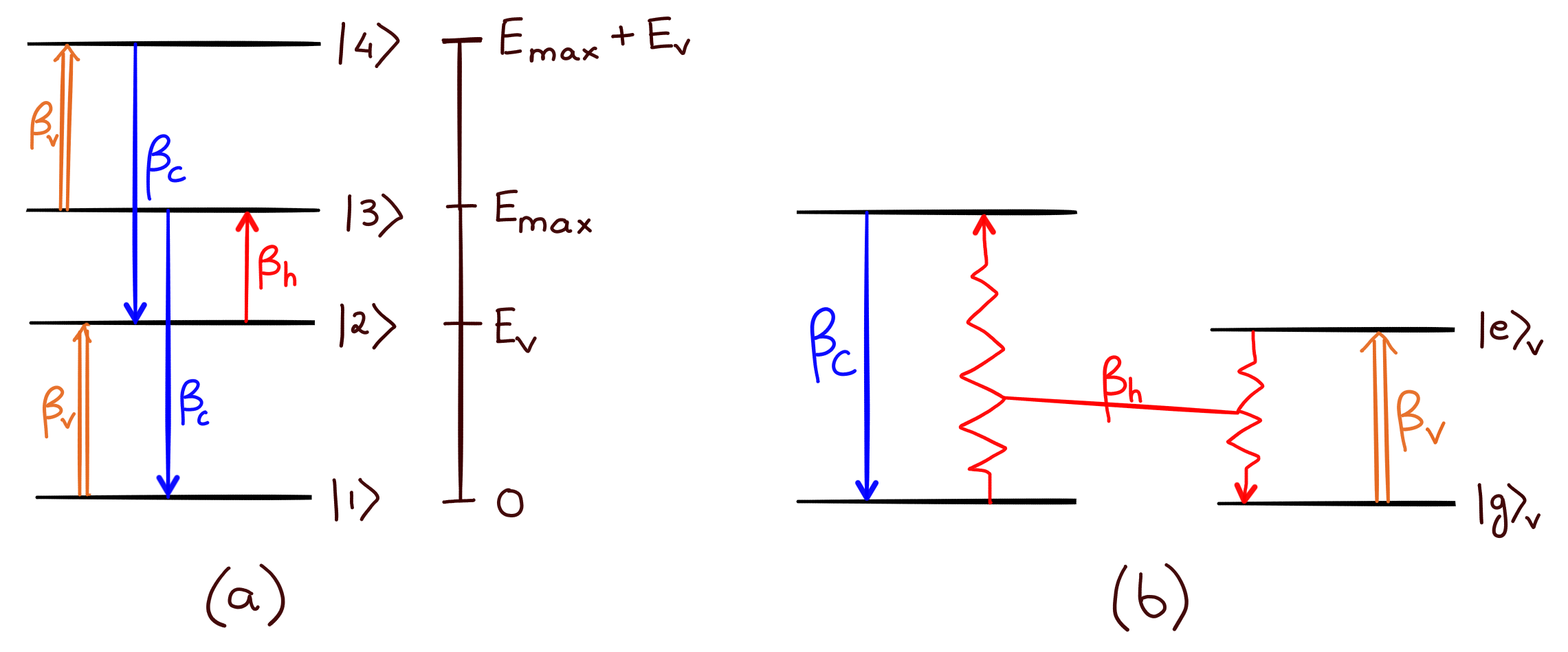}
\caption{Starting from the qutrit fridge, and adding a fourth level $\ket{4}$, the norm of the virtual qubit can be increased to $N_{\rm v}=1$, while 
maintaining the same bias $Z_{\rm v}$. This four-level fridge thus outperforms the qutrit fridge. (b) The four-level fridge viewed as a tensor product 
of the virtual qubit, now becoming a real qubit since $N_{\rm v}=1$, and a simpler thermal cycle. Note the coupling to the hot bath is now nonlocal, 
between the levels $\ket{0} \otimes \ket {e}_{\rm v}$ and $\ket{1} \otimes \ket{g}_{\rm v}$.  \label{improveNv}}
\end{figure}

For clarity, we illustrate the method starting from the qutrit fridge, that has a virtual qubit whose norm is strictly smaller than $1$. 
By adding a fourth level, we will achieve $N_{\rm v}=1$, while maintaining the bias. The fourth level is chosen specifically so that 
$E_4 = E_{\rm v} + E_{\rm max}$, and the transition $\Gamma_{2,4}$ is coupled to the cold bath (see Fig.~\ref{improveNv}(a)).
Hence by design, the new transition $\Gamma_{3,4}$ has the same energy gap $E_{\rm v}$ as the original virtual qubit $\Gamma_{1,2}$.
Furthermore, one can verify that both transitions possess the same virtual temperature. In fact one can identify two $3-$level fridge 
cycles at work in the new system, $\{ \ket{2} \rightarrow \ket{3} \rightarrow \ket{1} \}$ and $\{ \ket{4} \rightarrow \ket{2} \rightarrow \ket{3} \}$. 
Thus one could also connect $\Gamma_{3,4}$ to the external system that is to be cooled. Since the two transitions can be coupled at the same time to 
the external system, they both contribute to the virtual qubit. Thus, the norm of the (total) virtual qubit is obtained by summing the populations 
of each transition (virtual qubit). As the two transitions include all four levels, we find that $N_{\rm v}=1$.

Alternatively, one could view the four level machine as consisting of two real qubits, see Fig.~\ref{improveNv} (b). As one of these real qubits 
corresponds to the virtual qubit, it follows that its norm must be $N_{\rm v}=1$. We term this procedure the {\it virtual qubit amplification} of a 
single cycle machine. Next, we show explicitly how to perform the above construction starting from any $n$ level single cycle machine. This general 
virtual qubit amplification procedure requires the addition of $n-2$ additional levels. This is the most economical procedure possible, since the original 
$n$ level cycle contains $n-2$ levels which do not contribute to the virtual qubit.

\begin{figure}[t!]
\includegraphics[width=\linewidth]{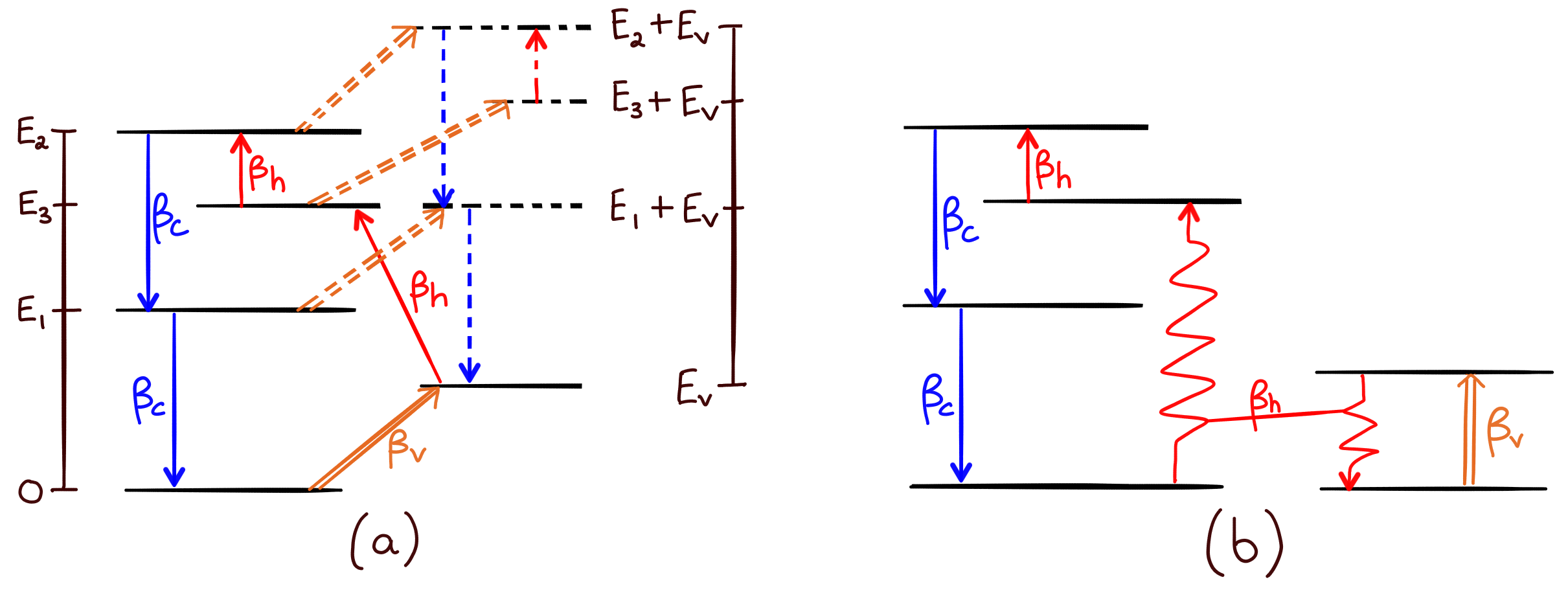}
\caption{(a) Starting from a 5 level fridge, and adding 3 levels (dashed lines), the norm of the virtual qubit can be boosted to $N_{\rm v}=1$ while maintaining the same bias $Z_{\rm v}$. (b) The resulting 8 level fridge can be viewed as a tensor product of a $4-$level cycle and the virtual qubit, which is now a real one since $N_{\rm v}=1$. \label{qubitrealization1}}
\end{figure}

The general construction works as follows. Consider a single $n$-level thermal cycle machine as described in Sec.~\ref{sec:single-cycle} : a 
set of $n$ levels with corresponding energies $E_{j}$ ($1\leq j \leq n$), subsequent $n-1$ transitions coupled to thermal baths at corresponding 
inverse temperatures $\beta_{j, {j+1}}$, and virtual qubit $\Gamma_{1, n}$, where $E_{n} - E_{1} = E_{\rm v}$. 
To amplify the virtual qubit, one now adds $n-2$ energy levels. Each new level is added in order to form a virtual qubit with each level of the 
original cycle except for the virtual qubit levels $\ket{1}$ and $\ket{n}$ (see Fig.~\ref{qubitrealization1}). The energy of the new 
levels must be chosen such that 
\begin{align}
	E_{{j+n-1}} = E_{j} + E_{\rm v}
\end{align}
where $j$ runs from $2$ to $n-1$. The corresponding thermal couplings are chosen in such a manner that the structure of the cycle from 
$j=n$ to $j=2n-2$ is identical to the structure from $j=1$ to $j=n-1$. Specifically, this means choosing 
\begin{align}
	\beta_{{j+n-2},{j+n-1}} &= \beta_{{j-1},j}. 
\end{align}
Following this procedure we finish with a final Hilbert space for the machine $\mathcal{H}$ with total dimension 
$n' \equiv {\rm dim}\mathcal{H} =2(n-1)$. One can verify that all the new virtual qubits ($\Gamma_{{1+j}, {n+j}}$) have the same virtual temperature 
$\beta_{\rm v}$ as the original virtual qubit $\Gamma_{1, n}$. None of these transitions share an energy level, i.e. they are mutually exclusive, and 
together they comprise all of the $2n-2$ levels present in the system. If every one of these transitions is connected together to the external system, 
then the effective virtual qubit reaches norm $N_{\rm v}=1$ as required. The inverse virtual temperature of the multi-cycle fridge can hence be expressed in terms of the total number of levels $n'$. For instance in the case of $n$ even, we have:
\begin{equation}\label{eq:multibetafin}
 \beta_{\rm v}^{(n')} = \beta_{\rm c} + (\beta_{\rm c} - \beta_{\rm h})\left(\frac{n'}{4} - \frac{1}{2} \right) \frac{E_{\rm max}}{E_{\rm v}}.
\end{equation}

Finally we note that, as in the simple case discussed above, the final machine can be viewed as a tensor product of an $n-1$ level cycle 
and the virtual qubit (which now becomes a real qubit since $N_{\rm v}=1$). In fact, this procedure also allows one to easily convert 
a fridge into a heat engine, and vice versa, as discussed in Appendix \ref{switching}. The virtual qubit amplification procedure is schematically 
depicted for the case of a $5-$level fridge cycle in Fig.~\ref{qubitrealization1}.

\section{Concatenated qutrit machines} \label{sec:concatenated}

As we commented previously, a different possibility for generalizing the simplest qutrit machine consists in concatenating several 
qutrit machines. Here we analyze this possibility by characterizing the virtual qubits achievable by concatenating $k$ qutrit machines (see Sec.~\ref{sec:qutrit}).

\begin{figure}[t!]
\includegraphics[width=\linewidth]{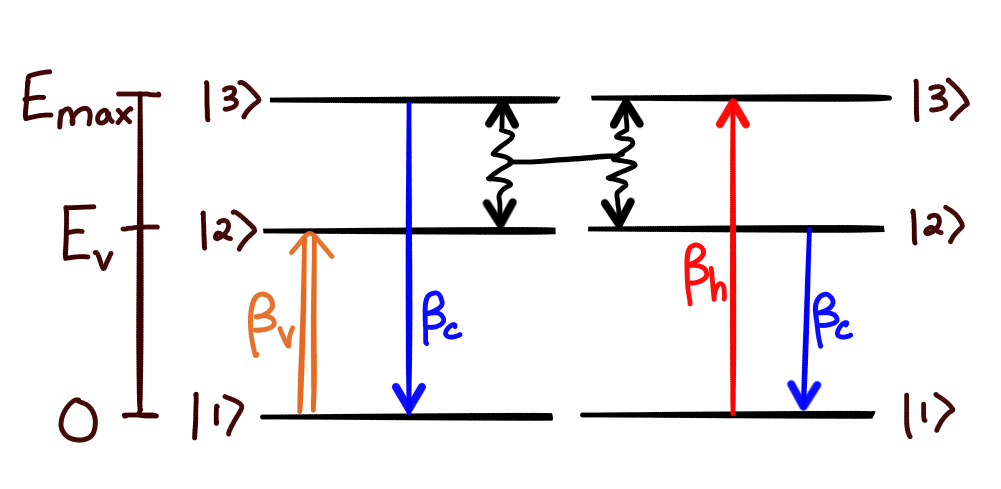}
\caption{By concatenating two qutrit machines, one obtains a better fridge, outperforming the simple qutrit fridge. Specifically, the new $6$-level 
machine consists now a qutrit fridge (left) which is boosted via the use of a qutrit heat engines (right). The role of this heat engine is to create 
an effectively hotter temperature (hotter than $T_{\rm h}$) in order to fuel the fridge. \label{2qutrit}}
\end{figure}

For simplicity we start with case of concatenating $k=2$ qutrit machines in order to obtain a better fridge. The coupling between the two qutrit 
machines can be achieved considering a simple swap Hamiltonian coupling the transitions $\Gamma_{2,3}^{(1)}$ and $\Gamma_{2,3}^{(2)}$:
\begin{align}
	H_{\rm int} &= g ( \ket{2,3}\bra{3,2} + {\rm h.c.} ),
\end{align}
as shown on Fig.~\ref{2qutrit}. Here the first qutrit machine represents the actual fridge while the second one works as a heat engine, replacing 
the hot bath on the transition $\Gamma_{2,3}^{(1)}$. This corresponds to coupling $\Gamma_{2,3}^{(1)}$ to an effective temperature which is hotter 
than \orange{the temperature of the hot bath (or equivalently inverse temperature lower than $\beta_{\rm h}$)}, resulting in a fridge 
\orange{with an improved} bias $Z_{\rm v}$.
Indeed the \orange{inverse} virtual temperature achieved by the concatenated qutrit machine is found to be 
\begin{align}\label{betavfridge2}
	\beta_{\rm v}^{(2)} = \beta_{\rm c} + (\beta_{\rm c}-\beta_{\rm h}) \frac{E_{\rm max}}{E_{\rm v}},
\end{align}
which is colder than the virtual temperature of the simple qutrit fridge (see Eq.~\eqref{betavfridge1}). Importantly, this enhancement 
has been achieved without modifying the value of $E_{\rm max}$, and considering the same temperatures $\beta_{\rm c}$ and $\beta_{\rm h}$ \orange{for the thermal 
baths}.
Details about calculations are given in Appendix \ref{qutrits}. 

\begin{figure}[t!]
\includegraphics[width=\linewidth]{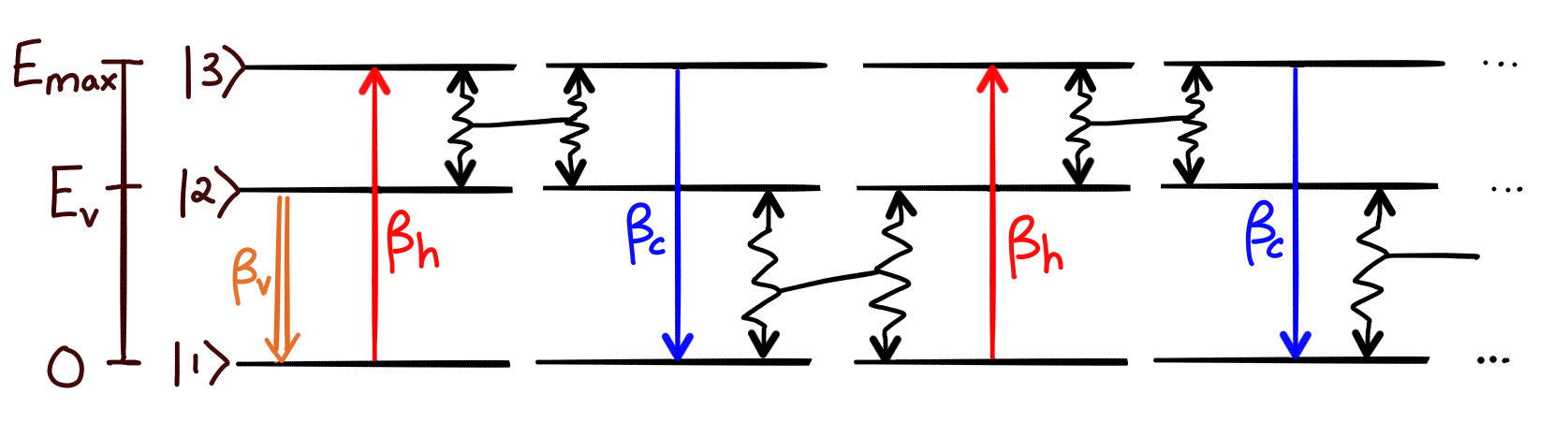}
\caption{Concatenating many qutrit machines.\label{nqutritlimit}}
\end{figure}

The process may now be iterated, replacing the coupling of $\Gamma_{2,3}^{(2)}$ to the cold bath $\beta_{\rm c}$ by a coupling to a third qutrit fridge, 
effectively at a temperature colder than $\beta_{\rm c}$, and so on, as sketched in Fig.~\ref{nqutritlimit}. In this manner one can construct a machine 
resulting of the concatenation of $k$ qutrit machines. Following calculations given in Appendix \ref{qutrits}, we obtain simple expressions 
for the virtual temperatures 
\begin{equation}\label{nqutrit}
	\beta_{\rm v}^{(k)} =  
		\begin{cases}
		\beta_{\rm c} + (\beta_{\rm c} - \beta_{\rm h})\frac{k}{2}\frac{E_{\rm max}}{E_{\rm v}} & \text{ if $k$ is even,} \\
		\beta_{\rm c} + (\beta_{\rm c} - \beta_{\rm h})\left( \frac{k+1}{2}\frac{E_{\rm max}}{E_{\rm v}} - 1 \right) & \text{ if $k$ is odd.}
		\end{cases}
\end{equation}
Again, we see that the virtual temperature approaches absolute zero as $k$ becomes large. Similarly for a concatenated heat engine, one can 
approach perfect inversion \orange{(see details in Appendix \ref{qutrits})}. 

\twocolumngrid

Note that the above expressions are similar to those obtained for the virtual temperature in the case of the single cycle machine. In particular 
setting $k=n-2$ we obtain exactly the same result. This correspondence can be intuitively understood via the following observations. First, the 
single qutrit machine is the same as a $3-$level cycle. Furthermore, the effect of replacing one of the thermal couplings in a qutrit machine by 
a coupling to an additional qutrit effectively replaces one thermal coupling by two, thus increasing the number of thermal interactions within 
the working cycle by one. For example, in the two qutrit fridge (Fig.~\ref{2qutrit}), the effective thermal cycle is
\begin{equation}
	\ket{22} \xrightarrow{\beta_{\rm c}} \ket{21} \xrightarrow{\beta_{\rm h}} \ket{23} \xrightarrow{H_{\rm int}} \ket{32} \xrightarrow{\beta_{\rm c}} \ket{12}.
\end{equation}
Although this is a cycle of length 5, the virtual temperature is only influenced by the 3 thermal couplings, \orange{because} the coupling on the 
degenerate transition $\ket{23}\leftrightarrow\ket{32}$ \orange{has zero energy gap (}see Eq.~\eqref{cyclebetav}). 
Since the thermal couplings are the same as those in the optimal $4-$level fridge single cycle, we get the same virtual temperature. By induction, the 
$k-$qutrit machine has the same $\beta_{\rm v}$ (and indeed the same thermal couplings within its working cycle) as the optimal $(k+2)-$level single cycle. 

\orange{Finally,} it is also important to discuss the behavior of the norm $N_{\rm v}$ of the virtual qubit in order to characterize the performance of the 
concatenated machine. Interestingly we find that $N_{\rm v} \rightarrow 1$ in the limit of large $k$.
This can be intuitively understood for the case of the concatenated heat engine, \orange{depicted in} Fig.~\ref{nqutritlimit}. As $k$ becomes large, 
the virtual temperature $\beta_{\rm v}$ approaches $-\infty$. Thus the population ratio $\frac{p_1}{p_2} \rightarrow 0$, implying that $p_1 \rightarrow 0$. 
However, since $\Gamma_{1,3}^{(1)}$ is coupled to a thermal bath at $\beta_{\rm h}$, the population ratio $\frac{p_3}{p_1}$  equals $e^{-\beta_{\rm h} E_{\rm max}}$, 
implying that $p_3 \rightarrow 0$. Thus in the limit $k \rightarrow \infty$, the state of the first qutrit approaches the pure state $\ket{2}\bra{2}$, 
and thus $N_{\rm v} = p_1 + p_2 \rightarrow 1$.
To understand the case of the fridge, consider in Fig.~\ref{nqutritlimit} that the machine begins with the second qutrit instead of the first one. 
This is now a fridge, where the virtual qubit is the transition $\Gamma_{2,3}^{(2)}$. By a similar analysis to the above, we find that the state of 
the qutrit approaches $\ket{2}\bra{2}$ in the limit $k \rightarrow \infty$, and thus $N_{\rm v} \rightarrow 1$.
It is instructive to observe that in both cases, the concatenation of qutrit machines takes the state of the original qutrit closer to the state 
where all of the population is in the middle level $\ket{2}\bra{2}$, which is both the ideal fridge with respect to $\Gamma_{2,3}$, and the ideal 
machine with respect to $\Gamma_{1,2}$.

Therefore we can conclude that, again, increasing the number of levels, or equivalently the dimension \orange{of the machine Hilbert space, 
$n \equiv {\rm dim}\mathcal{H} = 3^k$}, the performance is increased. Indeed, as $k$ increase, the virtual qubit bias approaches $Z_{\rm v}=1$ 
(or $Z_{\rm v}=-1$ for a heat engine), while \orange{its} norm becomes maximal, i.e. $N_{\rm v} \rightarrow 1$. However notice that \orange{in this case} 
the dimension of the machine grows rapidly. \orange{Indeed the inverse virtual temperature now grows only logarithmically with the total number of 
levels, $n$. For instance when $k$ is even we have: 
\begin{equation}
 \beta_{\rm v}^{(n)} = \beta_{\rm c} + (\beta_{\rm c} - \beta_{\rm h}) \left( \frac{\log_3 n}{2} \right) \frac{E_{\rm max}}{E_{\rm v}}
\end{equation}
to be compared with the multi-cycle fridge case in Eq.~\eqref{eq:multibetafin}.}

\orange{
\section{Third law} \label{sec:thirdlaw}

The above results show that when the dimension of the Hilbert space of the thermal machine tends to infinity, the virtual temperature can approach absolute zero even though the maximal 
energy gap which is coupled to a thermal bath is finite. Nevertheless, an important point is that, in all the constructions given, for any finite $n$, the lowest possible temperature is 
always strictly greater than zero. This can be directly seen from the expressions for the inverse virtual temperature of the optimal single-cycle machines, as given in Eq.~\eqref{betav} 
and Appendix \ref{AppOptTech}. Therefore any single-cycle fridge requires an infinite number of levels in order to cool to absolute zero.

Next, we notice that the lowest temperatures of any other mutli-cycle machine with different virtual qubits working in parallel can achieve is bounded 
by the temperature achieved in any of these cycles. This follows from the fact that the effect of multiple cycles on the virtual qubit can be 
decomposed as a sum of the effect of each individual cycle. Thus, the bound on the temperature we derive for single-cycle $n$ level machines holds 
for general machines with $n$ levels.}

Therefore we obtain a statement of the third law in terms of Hilbert space dimension. In particular, from \eqref{swapeffect} we see that the bias (and therefore temperature) and norm of the virtual qubit determine to what temperature an external object can be bought to in a single (or multiple) cycles of a thermal machine. The fact that the virtual temperature only approaches zero as the dimension of the thermal machine approaches infinity shows that an brining an external object to absolute zero requires a machine with an infinite number of levels. This is static version of the third law, complementary to previous statements \cite{Masanes,amikam}, stated in terms of number of steps, time, or energy required in order to reach absolute zero. 

Finally, we note that in the case of the multi-cycle machine, since the norm of the virtual qubit is unity, in a single swap operation the external object is bought to exactly the temperature of the virtual qubit. Thus, using a machine of Hilbert space dimension $n$, we can cool an external object to the inverse temperature \eqref{eq:multibetafin}, which corresponds asymptotically to the scaling
\begin{equation}
T_\mathrm{s} \sim \frac{1}{n}
\end{equation}
\textit{i.e.}~the temperature scales inversely with the Hilbert space dimension. 

\section{Statics vs dynamics for single-cycle machines}\label{sec:dynamics}

So far, we have discussed improving the {\it static} configuration of the thermal machine by increasing its dimension. This analysis characterizes 
the task of cooling (or heating) an external system via a single swap, a so-called {\it single shot} thermodynamic operation. However, more 
generally we are interested in continuously cooling the external system, as the latter is unavoidably in contact with its own 
environment, and thus requires repeated swaps with the virtual qubit in order to maintain the cooling (or heating) effect.

As we have seen in Sec.~\ref{sec:primitive}, after a single swap between the virtual qubit and the external system, the bias of the 
virtual qubit is switched with that of the external system. Thus the virtual qubit needs to be ``reset'' before the next interaction is possible, 
an operation which should require some time to be performed, and hence introduces limitations on the power of the machines. This 
``time of reset'' depends in general on the thermalization model, which forces us to go beyond purely static considerations. To illustrate this 
point we will discuss here the dynamics of the single-cycle refrigerators.

Intuitively one may expect the time of reset of the virtual qubit increases as the number of levels in the cycle increases, i.e. the larger the cycle of the machine, the longer it takes the machine to perform the series of jumps reinitializing it. This introduces 
the following tradeoff. Previously we saw that machines with longer cycles were able to achieve lower temperatures for a single swap. 
However, they would also take longer to reset. Therefore in order to engineer a good fridge, one could consider (i) a high dimensional fridge 
(i.e. a long cycle) achieving low temperatures at slower rate, or (ii) a low-dimensional fridge achieving not as low temperatures, but at a faster rate. 

In order to find out which regime is better, we consider single-cycle fridges coupled to thermal baths, as modelled by a Markovian master equation. Since the thermalization occurs here only on transitions, the specific details of the model are not crucial, and all models (either simple heuristic ones \cite{linden10} or those derived explicitly by microscopic derivations \cite{breuer}) lead to the same qualitative conclusions.

\begin{figure}[t!]
\includegraphics[width=\linewidth]{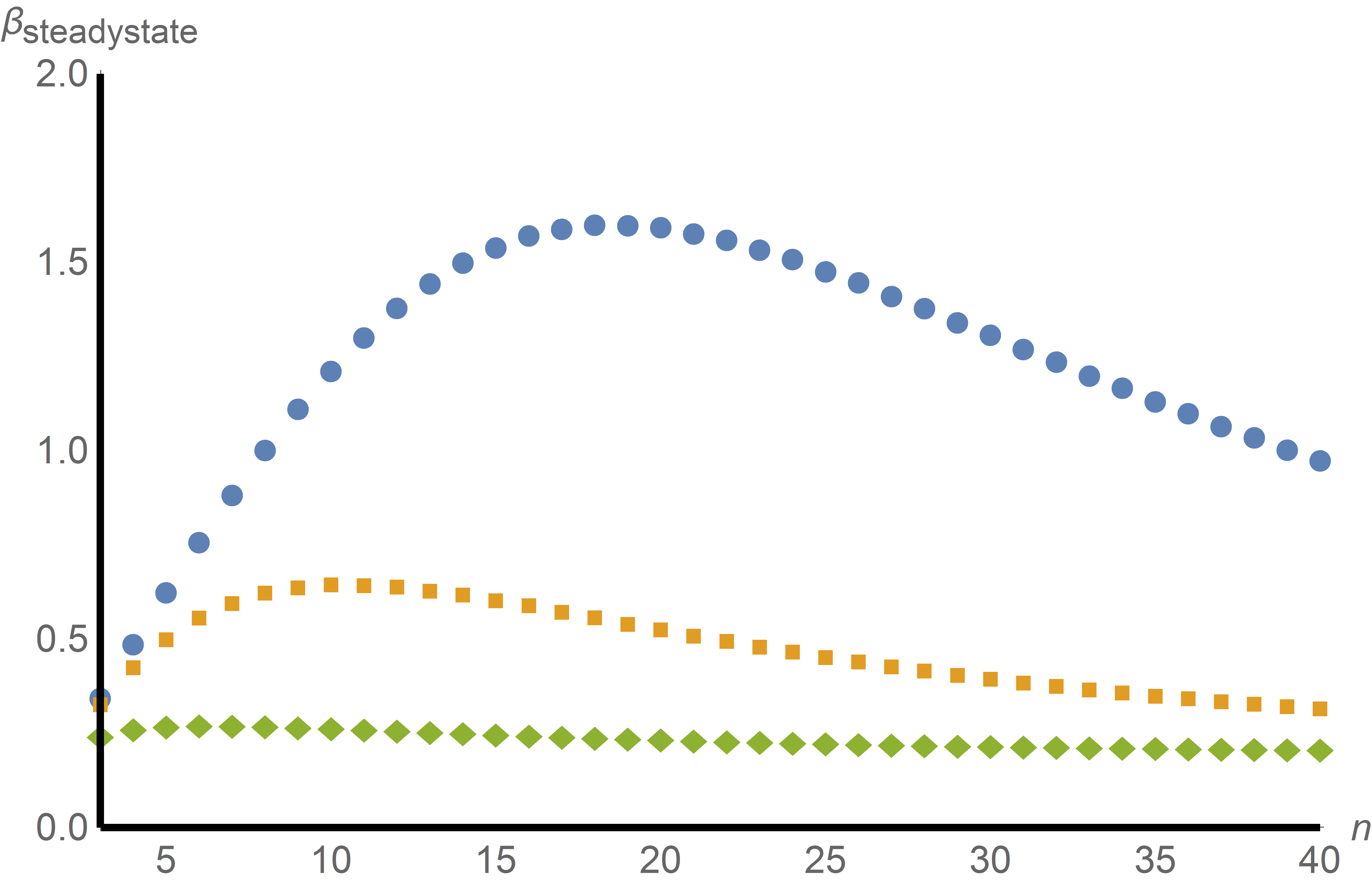}
\caption{Relationship between the steady-state virtual temperature and the length of the cycle. We consider various equilibration timescales, 
$\tau_{\rm s} = 1$ (green, diamond), $\tau_{\rm s} = 10$ (orange, square) and $\tau_{\rm s} = 100$ (blue, dot). All other parameters are kept fixed: timescale of 
all thermal couplings of the cycle $\tau_\beta =1$, bath temperatures $\beta_{\rm h} = 0.05$, $\beta_{\rm c} = 0.2$, and energies $E_{\rm max} = 2$, and $E_{\rm v} = 1$ 
(as in Fig.~\ref{mainres}).\label{betavsN}}
\end{figure}

We find that the relevant parameter is timescale at which the external system interacts with its environment $\tau_{\rm s}$. If this timescale is short, then the fridge has little time to `reset' the virtual qubit. Therefore a shorter cycle, that resets quickly, is optimal in this case. If on the contrary the system timescale is long, there is more time available in order to reset the virtual qubit. Thus a longer cycle, providing lower temperatures, is preferable. This trade-off is illustrated in Fig.~\ref{betavsN}. 

\twocolumngrid

\begin{figure}[h!]
\includegraphics[width=\linewidth]{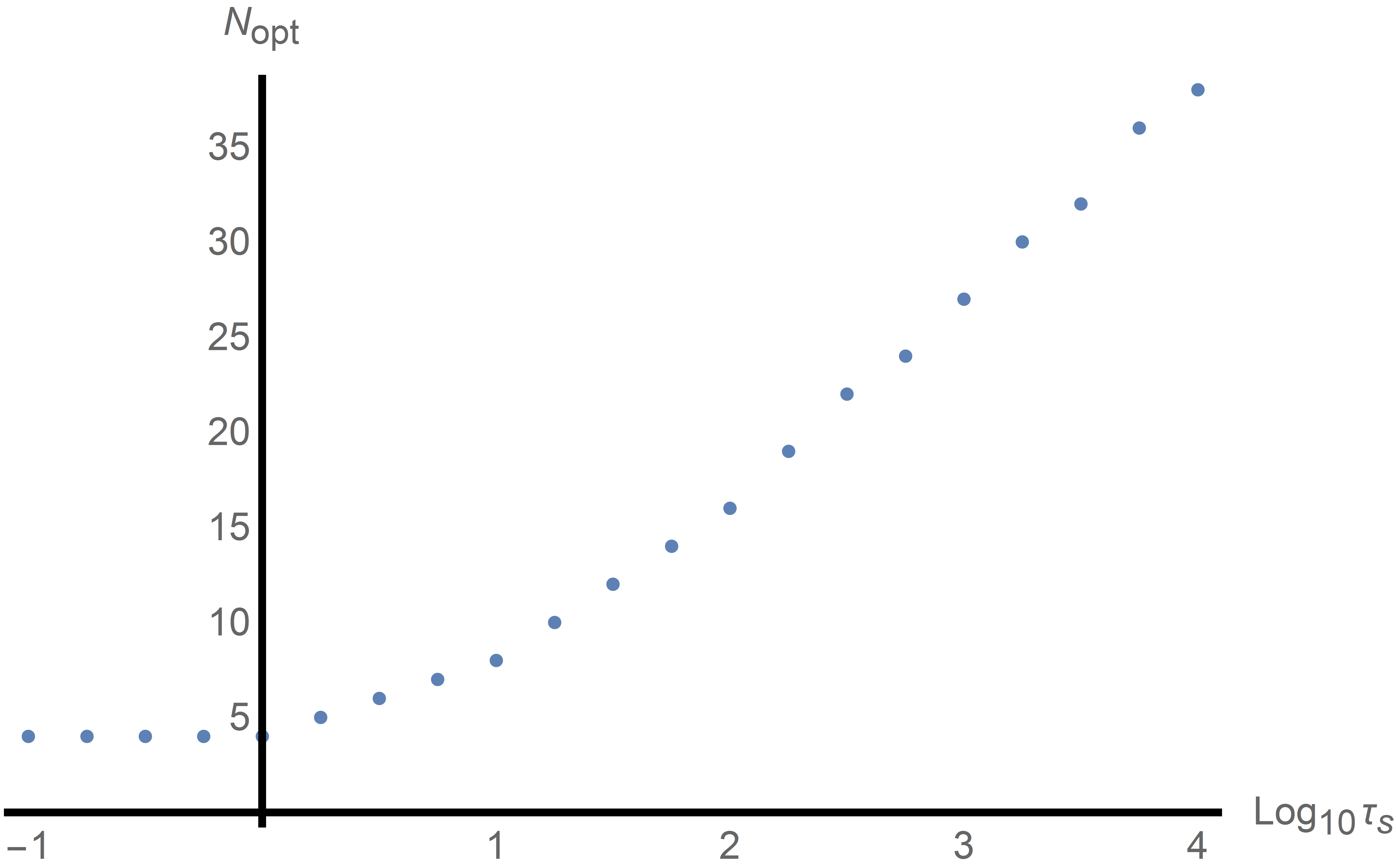}
\caption{Length of the optimal cycle versus equilibration timescale $\tau_{\rm s}$. Other parameters are the same as in Fig.~\ref{betavsN}.\label{optnvst}}
\end{figure}

We also observe from Fig.~\ref{betavsN} that, for given timescale $\tau_{\rm s}$, there is an optimal length of the cycle. In Fig.~\ref{optnvst}, 
we plot the optimal length of the cycle for different timescales. The optimal length appears to be logarithmic with respect to $\tau_{\rm s}$. However, 
for fast timescales, we observe that the optimal cycle has length 4. This suggests that the simplest qutrit machine is always outperformed in this 
regime. 

\twocolumngrid

\section{Discussion/Conclusion} \label{sec:conclusions}

We discussed the performance of quantum absorption thermal machines, in particular with respect to the size of the machine. Specifically, we 
considered several designs of machines with $n$ levels and described the static properties of the machine, in particular the range of available 
virtual qubits, which characterizes the fundamental limit of the machine. Notably, as $n$ increases, a larger range of virtual temperatures 
becomes available, showing that a machine with $n+1$ levels can outperform a machine with $n$ levels. Moreover, in order to achieve virtual 
qubits with perfect bias (i.e. achieving a virtual qubit at zero temperature, or with complete population inversion), the required number of 
levels $n$ diverges. This can be viewed as a statement of the third law, complementary to previous ones. Usually stated in terms of number of 
steps, time, or energy required in order to reach absolute zero temperature, we obtain here a statement of the third law in terms of Hilbert 
space dimension: reaching absolute zero requires infinite dimension.

Moreover, we also discussed machines with multiple cycles running in parallel. Here performance is increased, as the norm of the virtual qubit 
can be brought to one, i.e. the virtual qubit becomes a real one. Finally, similar performance is achieved for a design based on the concatenation 
of the simplest qutrit machine. While generally suboptimal in terms of performance, this design gives nevertheless a more intuitive picture and 
may be more amenable to implementations, as the couplings are simpler.

An outstanding question left open here concerns the performance of machine where multiple cycles run in parallel. In particular, it would be 
interesting to understand how to design the most effective machine, given a fixed number of levels (as well as constraints on the energy and 
temperatures). One may expect that the time necessary to reset the machine is considerably decreased, providing potentially a strong advantage 
over single-cycle machines.

\section{Acknowledgements} We thank Sandu Popescu for discussions. We acknowledge financial support from the European project ERC-AD NLST, the 
Swiss National Science Foundation (grant PP00P2\_138917 and QSIT). \orange{G. M. acknowledges funding from MINECO (grants FIS2014-52486-R and 
BES-2012-054025). This work has been supported by the COST Action MP1209 ``Thermodynamics in the quantum regime''.}

\onecolumngrid
\newpage
\section*{APPENDICES}
\appendix

\twocolumngrid

\section{The swap operation as the primitive for thermodynamic operations.}\label{AppSwap}

This appendix elaborates on the swap as the primitive operation of quantum thermal machines. Consider a setup involving a real qubit of energy $E_{\rm v}$, the system, with bias $Z_{\rm s}$.

In order to modify the bias (e.g. to cool the system), the system now interacts with a virtual qubit (i.e. a pair of levels $\{i,j\}$ within the machine) which has the same energy gap as the system, i.e. $E_{\rm v}=E_j - E_i $. The  energy-conserving ``swap" interaction is described by a unitary
\begin{align}
	U = \mathbb{I}_{\rm sv} &- \ket{i,0}_{\rm sv}\!\bra{i,0} - \ket{j,1}_{\rm sv}\!\bra{j,1} \nonumber\\ &+ \ket{i,1}_{\rm sv}\!\bra{j,0} + \ket{j,0}_{\rm sv}\!\bra{i,1},
\end{align}
where $\ket{0}_{\rm s}$ and $\ket{1}_{\rm s}$ denote the ground and excited states of the system.

The effect of the swap upon two real qubits would be to swap the states of the qubits for one another (assuming the initial state as diagonal and uncorrelated). However, this is not the case for one real and one virtual qubit, as we show presently.

We assume that the real qubit begins in a diagonal state. If one labels the populations of the initial state in the ground and excited levels of the system as $p_0$ and $p_1$, then using the definition of the bias, $Z = p_0 - p_1$, its initial state is
\begin{equation}
	\rho_{\rm s} = \frac{1+Z_{\rm s}}{2} \ket{0}_{\rm s}\!\bra{0} + \frac{1-Z_{\rm s}}{2} \ket{1}_{\rm s}\!\bra{1}.
\end{equation}

For the virtual qubit, the sum of the populations is not $1$ in general, i.e. $N_{\rm v} = p_i + p_j < 1$. Assuming that the state is block diagonal (w.r.t. the virtual qubit),
\begin{align}
	\rho_{\rm v} = N_{\rm v} \left( \frac{1 + Z_{\rm v}}{2} \ket{i}_{\rm v}\!\bra{i} + \frac{1-Z_{\rm v}}{2} \ket{j}_{\rm v}\!\bra{j} \right) \nonumber \\
	+ (1 - N_{\rm v}) \rho^\prime_{\rm v},
\end{align}
where $\rho^\prime_{\rm v}$ is an arbitrary (normalized) state of the remaining levels in the machine.

After applying $U$, the final state of the system and the machine containing the virtual qubit is
\onecolumngrid
\begin{align}
	U \rho_{\rm s} \otimes \rho_{\rm v} U^\dagger &= \left( \frac{1+Z_{\rm s}}{2} \right) N_{\rm v} \left( \frac{1+Z_{\rm v}}{2} \right) \ket{00}_{\rm sv}\!\!\bra{00} + \left( \frac{1-Z_{\rm s}}{2} \right) N_{\rm v} \left( \frac{1+Z_{\rm v}}{2} \right) \ket{01}_{\rm sv}\!\!\bra{01} \nonumber\\
	&+ \left( \frac{1+Z_{\rm s}}{2} \right) N_{\rm v} \left( \frac{1-Z_{\rm v}}{2} \right) \ket{10}_{\rm sv}\!\!\bra{10} + \left( \frac{1-Z_{\rm s}}{2} \right) N_{\rm v} \left( \frac{1-Z_{\rm v}}{2} \right) \ket{11}_{\rm sv}\!\!\bra{11} + (1-N_{\rm v}) \rho_{\rm s} \otimes \rho^\prime_{\rm v},
\end{align}
from which the final reduced state of the system is
\begin{align}
	\rho_{\rm s}^f &= \left[ N_{\rm v} \left( \frac{1+Z_{\rm v}}{2} \right) + (1-N_{\rm v}) \left( \frac{1+Z_{\rm s}}{2} \right) \right] \ket{0}_{\rm s}\!\!\bra{0} + \left[ N_{\rm v} \left( \frac{1-Z_{\rm v}}{2} \right) + (1-N_{\rm v}) \left( \frac{1-Z_{\rm s}}{2} \right) \right] \ket{1}_{\rm s}\!\!\bra{1}.
\end{align}

At the end of the protocol, the bias of the real qubit has been modified to
\begin{equation}\label{appbiaschange}
	Z_{\rm s}^\prime = N_{\rm v} Z_{\rm v} + (1-N_{\rm v}) Z_{\rm s} \;\;\;\;\;\Longrightarrow\;\;\;\;\; \Delta Z_{\rm s} = Z_{\rm s}^\prime - Z_{\rm s} = N_{\rm v} \left( Z_{\rm v} - Z_{\rm s} \right).
\end{equation}

\section{Optimal single cycle machines}\label{AppOptTech}
\twocolumngrid

We prove optimality of the single cycle machine discussed in Section \ref{sec:single-cycle} of the main text. While there are several ways in which performance could be discussed, we are mainly concerned here with optimality under the swap operation \eqref{appbiaschange}. That is, which machine achieves the largest change in the bias of the system acted upon.

Consider a machine with $n$ levels and a single cycle (of length $n$). All transitions must be coupled to available temperatures, namely
\begin{equation}\label{constraintbetaapp}
	\beta_{\rm h} \leq \beta_{j,j+1} \leq \beta_{\rm c}.
\end{equation}
Note that intermediate temperatures can be obtained by coupling to both baths at $\beta_{\rm c}$ and $\beta_{\rm h}$. Furthermore, the energy gaps of the transitions are bounded,
\begin{equation}\label{constrainttransition}
	- E_{\rm max} \leq \Delta E_{j,j+1} \leq E_{\rm max}.
\end{equation}

The cycle approaches here a diagonal steady state, as every level is interacting with a thermal bath. The ratio of populations of every transition matches the temperature of the bath it is coupled to,
\begin{equation}
	\frac{p_{j+1}}{p_{j}} = e^{-\beta_{j,j+1} \Delta E_{j,j+1}} \quad \text{for  } 1 \leq j \leq n-1.
\end{equation}
Together with the normalization condition $\sum_j p_{j} = 1$, this completely determines the steady state. The virtual temperature $\beta_{\rm v}$ is given by
\begin{align}
	e^{-\beta_{\rm v} E_{\rm v}} &= \frac{p_{n}}{p_{1}} = \frac{p_{n}}{p_{n-1}} \frac{p_{n-1}}{p_{n-2}}...\frac{p_{2}}{p_{1}}, \\ 
	\therefore \beta_{\rm v} E_{\rm v} &= \sum_{j=1}^{n-1} \beta_{j,j+1} \Delta E_{j,j+1}. \label{cyclebetavapp}
\end{align}
Similarly, the norm $N_{\rm v}$ is found to be
\begin{align}
	N_{\rm v} &= \frac{p_{1}+p_{n}}{p_{1} \left( 1 + \frac{p_{2}}{p_{1}} + \frac{p_{3}}{p_{1}} + ... \right)} \\
	&=\left( \frac{1 + e^{-\beta_{\rm v} E_{\rm v}}}{1 + \sum_{j=1}^{n-1} \prod_{k=1}^{k=j} e^{-\beta_{k,k+1} \Delta E_{k,k+1}}} \right).\label{cycleNvapp}
\end{align}

We proceed to determine the unique $n$ level single cycle that \emph{minimizes the ratios of the population of every level ${j}$ in the cycle with respect to one of the levels of the virtual qubit.} This is then proven to be the optimal cycle. For clarity we detail the proof for the case of the fridge, i.e. we minimize the ratios w.r.t. the ground state of the virtual qubit. The proof for the heat engine is similar.

Consider the population ratio
\begin{align}
	\frac{p_{j}}{p_{1}} &= \prod_{k=1}^{j-1} e^{-\beta_{k,k+1} \Delta E_{k,k+1}} \\
		&= \exp \left[ -\sum_{k=1}^{j-1} \beta_{k,k+1} \Delta E_{k,k+1} \right].
\end{align}
To minimize this ratio, one should maximize the summation above. Regardless of the values of any energy gap $\Delta E$, maximizing the sum requires picking the highest possible temperature $\beta_{\rm c}$ if the energy gap is positive, and the smallest possible temperature $\beta_{\rm h}$ if the energy gap is negative. Thus one can collect together the positive and negative energy gaps to simplify the expression. Labeling the sum of the positive energy gaps as $Q_+^j$ and the sum of the negative ones as $Q_-^j$, we obtain
\begin{equation}\label{ratiogeneral}
	\frac{p_{j}}{p_{1}} = \exp \left[ - \left( \beta_{\rm c} Q_+^j + \beta_{\rm h} Q_-^j \right) \right].
\end{equation}
In addition, we have the consistency relation
\begin{equation}\label{Qconsistency}
	Q_+^j + Q_-^j = \Delta E_{1,j} = \sum_{k=1}^{j-1} \Delta E_{k,k+1},
\end{equation}
which leads to 
\begin{equation}\label{twostep}
	\frac{p_{j}}{p_{1}} = \exp \left[ - \beta_{\rm h} \Delta E_{1,j} - \left( \beta_{\rm c} - \beta_{\rm h} \right) Q_+^j \right].
\end{equation}

We proceed to minimize the ratio in two steps. First we find the optimum $Q_+^j$ for a fixed $\Delta E_{1,j}$, followed by optimizing over $\Delta E_{1,j}$.

For a fixed energy gap $\Delta E_{1,j}$, the minimum Gibbs ratio is achieved when $Q_+^j$ is as large as possible (since $\beta_{\rm c}-\beta_{\rm h} >0$). Recall that $Q_+^j$ is the sum of positive transitions in the cycle from $1$ to $j$, each of which are bounded by $E_{\rm max}$. Also the number of transitions at $E_{\rm max}$ between $1$ to $j$ is limited by the consistency relation \eqref{Qconsistency}. Optimizing for $Q_+^j$ subject to these constraints results in values for the sizes and number of transition in the cycle from $1$ to $j$ as summarized in the Table \ref{transitionsizearbitrary}, for a fixed $\Delta E_{1,j} = m E_{\rm max} + \delta_j$ (where $m = \Delta E_{1,j} \mod E_{\rm max}$).

In spite of the dependence on the optimum current $Q_+^j$ upon the relative parities of $j$ and $m$, it is straightforward to verify that the optimum $Q_+^j$ increases monotonically w.r.t. $\Delta E_{1,j}$. Thus to complete the minimization of \eqref{twostep}, one has to maximize $\Delta E_{1,j}$. This proceeds in an analogous manner to the optimization of $Q_+^j$, with the major difference being that $\Delta E_{1,j}$ must be chosen keeping in mind the consistency condition for the energy gap of the virtual qubit \eqref{virtualgap}. The result is summarized in Table \ref{transitionsizeintermediate}, for the $n$ level cycle.

This completes the optimization of the ratio $p_{j}/p_{1}$. From Table \ref {transitionsizeintermediate} we see that there is a unique construction of the $n$ level cycle that simultaneously fulfils the optimization criteria for all $j$: for all $j\leq n/2$ fix all of the transitions to be $+E_{\rm max}$, next fix a transition to be $E_{\rm v}$ or $-(E_{\rm max} - E_{\rm v})$, depending on the parity of $n$, and continue with all the remaining transitions fixed to be $-E_{\rm max}$.

\onecolumngrid
\begin{center}
\begin{table}[b]
\renewcommand{\arraystretch}{2.0}
\begin{tabular}{| c | c | c | c | c || c | c |}
	\hline No. of transitions & $+E_{\rm max}$ & $+\delta_j$ & $-(E_{\rm max} - \delta_j)$ & $-E_{\rm max}$ & $Q_+$ & $Q_-$ \\
	\hline if $j$ and $m$ are both even or odd & $\frac{j+m}{2} - 1$ & $1$ & $0$ & $\frac{j-m}{2} - 1$ & $\left( \frac{j+m}{2} - 1 \right) E_{\rm max} + \delta_j$ & $-\left( \frac{j-m}{2} - 1 \right) E_{\rm max}$ \\
	\hline if $j$ and $m$ are of opposite parity & $\frac{j+m-1}{2}$ & $0$ & $1$ & $\frac{j-m-3}{2}$ & $\left( \frac{j+m-1}{2} \right) E_{\rm max}$ & $-\left( \frac{j-m-1}{2} \right) E_{\rm max} + \delta_j$ \\
	\hline
\end{tabular}
\caption{Transition number and size, and heat currents, to maximize the heat current $Q_+^j$ associated to an arbitrary level $j$ w.r.t. the first energy level, within a thermal cycle.}
\label{transitionsizearbitrary}
\end{table}
\end{center}

\begin{center}
\begin{table}[t]
\renewcommand{\arraystretch}{2.0}
\begin{tabular}{| c | c | c | c | c || c | c || c |}
	\hline No. of transitions & $+E_{\rm max}$ & $+E_{\rm v}$ & $-(E_{\rm max} - E_{\rm v})$ & $-E_{\rm max}$ & $Q^j_+$ & $Q^j_-$ & $\Delta E_{1,j}$ \\
	\hline if $j\leq \frac{n}{2}$ & $j-1$ & $0$ & $0$ & $0$ & $(j-1)  E_{\rm max}$ & $0$ & $(j-1) E_{\rm max}$ \\
	\hline $j>\frac{n}{2}$, $n$ even & $\frac{n}{2} - 1$ & $1$ & $0$ & $j - \frac{n}{2}$ & $\left( \frac{n}{2} -1 \right) E_{\rm max} + E_{\rm v}$ & $-\left( j- \frac{n}{2} \right) E_{\rm max}$ & $(n-j-1)E_{\rm max} + E_{\rm v}$ \\
	\hline $j>\frac{n}{2}$, $n$ odd & $\frac{n-1}{2}$ & $0$ & $1$ & $j - \frac{n+1}{2}$ & $\left( \frac{n-1}{2} \right) E_{\rm max}$ & $-\left( j - \frac{n-1}{2} \right) E_{\rm max} + E_{\rm v}$ & $(n-j-1) E_{\rm max} + E_{\rm v}$ \\
	\hline
\end{tabular}
\caption{Transition number and size, and heat currents, to minimize the Gibbs ratio of an arbitrary level $j$ w.r.t. the first energy level, within a thermal cycle.}
\label{transitionsizeintermediate}
\end{table}
\end{center}
\twocolumngrid

Finally, connecting all $+ve$ transitions to $\beta_{\rm c}$ and $-ve$ transitions to $\beta_{\rm h}$, one arrives at the optimal $n$ level cycle fridge, schematically depicted in Fig \ref{optimalcycle}. If we instead minimize the ratios of populations to the excited state of the virtual qubit ($p_{j}/p_{n}$), we obtain the optimal $n$ level cycle engine, which has the same arrangement of energy levels as the fridge, with only the temperatures swapped, $\beta_{\rm c} \leftrightarrow \beta_{\rm h}$.

For completeness, we present below the virtual temperatures $\beta_{\rm v}^{(n)}$ and norms $N_{\rm v}^{(n)}$ achieved by the optimal $n$ level cycle fridge and engine.

\onecolumngrid
\begin{center}
\begin{table}[h]
\renewcommand{\arraystretch}{2.0}
\begin{tabular}{| c | c | c |}
	\hline $\beta_{v}^{(n)} E_{\rm v}$ & $n$ even & $n$ odd \\
	\hline Fridge & $\;\;\; \beta_{\rm c} E_{\rm v} + \left( \beta_{\rm c} - \beta_{\rm h} \right) \left( \frac{n}{2} -1 \right) E_{\rm max} \;\;\;$ & $\;\;\; \beta_{\rm c} E_{\rm v} + \left( \beta_{\rm c} - \beta_{\rm h} \right) \left[ \left( \frac{n}{2} - \frac{1}{2} \right) E_{\rm max} - E_{\rm v} \right] \;\;\;$ \\
	\hline Engine & $\;\;\; \beta_{\rm h} E_{\rm v} - \left( \beta_{\rm c} - \beta_{\rm h} \right) \left( \frac{n}{2} -1 \right) E_{\rm max} \;\;\;$ & $\;\;\; \beta_{\rm h} E_{\rm v} - \left( \beta_{\rm c} - \beta_{\rm h} \right) \left[ \left( \frac{n}{2} - \frac{1}{2} \right) E_{\rm max} - E_{\rm v} \right] \;\;\;$ \\
	\hline
\end{tabular}
\caption{Optimal virtual temperatures of a thermal cycle of length $n$.}
\label{optimumbetav}
\end{table}
\renewcommand{\arraystretch}{2.0}
\begin{table}[h]
\begin{tabular}{| c | c |}
	\hline $N_{\rm v}^{(n)}$ & $n$-level optimal fridge cycle \\
	\hline $n_{even}$ & $\;\;\; \left( 1 + e^{-\beta_{\rm v}^{(n)} E_{\rm v}} \right) \left[ \left( 1 - e^{-\beta_{\rm c} E_{\rm max}} \right)^{-1} \left( 1 - e^{-\frac{n}{2} \beta_{\rm c} E_{\rm max}} \right) + e^{-\beta_{\rm v}^{(n)} E_{\rm v}} \left( 1 - e^{-\beta_{\rm h} E_{\rm max}} \right)^{-1} \left( 1 - e^{-\frac{n}{2} \beta_{\rm h} E_{\rm max}} \right) \right]^{-1} \;\;\;$ \\
	\hline $n_{odd}$ & $\;\;\; \left( 1 + e^{-\beta_{\rm v}^{(n)} E_{\rm v}} \right) \left[ \left( 1 - e^{-\beta_{\rm c} E_{\rm max}} \right)^{-1} \left( 1 - e^{-\left( \frac{n+1}{2} \right) \beta_{\rm c} E_{\rm max}} \right) + e^{-\beta_{\rm v}^{(n)} E_{\rm v}} \left( 1 - e^{-\beta_{\rm h} E_{\rm max}} \right)^{-1} \left( 1 - e^{-\left( \frac{n-1}{2} \right) \beta_{\rm h} E_{\rm max}} \right) \right]^{-1} \;\;\;$ \\
	\hline & $n$-level optimal engine cycle \\
	\hline $n_{even}$ & $\;\;\; \left( 1 + e^{+\beta_{\rm v}^{(n)} E_{\rm v}} \right) \left[ \left( 1 - e^{-\beta_{\rm c} E_{\rm max}} \right)^{-1} \left( 1 - e^{-\frac{n}{2} \beta_{\rm c} E_{\rm max}} \right) + e^{+\beta_{\rm v}^{(n)} E_{\rm v}} \left( 1 - e^{-\beta_{\rm h} E_{\rm max}} \right)^{-1} \left( 1 - e^{-\frac{n}{2} \beta_{\rm h} E_{\rm max}} \right) \right]^{-1} \;\;\;$ \\
	\hline $n_{odd}$ & $\;\;\; \left( 1 + e^{+\beta_{\rm v}^{(n)} E_{\rm v}} \right) \left[ \left( 1 - e^{-\beta_{\rm c} E_{\rm max}} \right)^{-1} \left( 1 - e^{-\left( \frac{n-1}{2} \right) \beta_{\rm c} E_{\rm max}} \right) + e^{+\beta_{\rm v}^{(n)} E_{\rm v}} \left( 1 - e^{-\beta_{\rm h} E_{\rm max}} \right)^{-1} \left( 1 - e^{-\left( \frac{n+1}{2} \right) \beta_{\rm h} E_{\rm max}} \right) \right]^{-1} \;\;\;$ \\
	\hline
\end{tabular}
\caption{Norm $N_{\rm v}$ of the optimal $n-$level thermal cycle, in terms of the virtual temperature $\beta_{\rm v}^{(n)}$.}
\label{optimalnorms}
\end{table}
\end{center}
\twocolumngrid

\subsection{Characterizations of optimality for single-cycle machines.}

Here we demonstrate useful properties of the optimal $n$ level cycle, in particular that it achieves the largest change in the bias of an external qubit under the swap operation.

Recall the technical definition in Appendix \ref{AppOptTech}, that the optimal cycle is the unique cycle (fridge) that minimizes the ratios of every single population to the ground state of the virtual qubit $p_{1}$. In particular, this includes the Gibbs ratio of the virtual qubit itself, $p_{n}/p_{1}$, and thus the optimal cycle \emph{maximizes the bias $Z_{\rm v}$}. In addition, using the normalization of the cycle $\sum_j p_{j} = 1$, one can express the norm of the virtual qubit in the useful form
\begin{align}\label{appcycleNv}
	N_{\rm v}&= \left( \frac{1 + e^{-\beta_{\rm v} E_{\rm v}}}{1 + \sum_{j=2}^{n} p_{j}/p_{n}} \right).
\end{align}

Since the optimal cycle is the unique cycle that minimizes the denominator above, in particular it does so for the case that $\beta_{\rm v}$ is the optimal temperature (corresponding to the optimum bias $Z_{\rm v}$), thus the optimal cycle \emph{achieves the highest norm $N_{\rm v}$ given the maximum bias $Z_{\rm v}$}.

Expressing the population of the ground state of the virtual qubit as
\begin{equation}\label{optp1}
	p_{1} = \frac{1}{1 + \sum_{j=2}^{n} p_{j}/p_{n}},
\end{equation}
it is clear that the optimal cycle also \emph{maximizes the population $p_{1}$}, which is equivalently the \emph{maximal value of $N_{\rm v} (1+Z_{\rm v})$}.

Since the optimal cycle both maximizes $p_{1}$ and minimizes $p_{n}/p_{1}$, we may conclude that it \emph{maximizes the difference between the populations}
\begin{equation}\label{optNvZv}
	p_{1} - p_{n} = N_{\rm v} Z_{\rm v} = p_{1} \left( 1 - \frac{p_{n}}{p_{1}} \right).
\end{equation}
	
Equivalently, in the case of the engine, the optimal $n$ level cycle:
\begin{itemize}
	\item minimizes $Z_{\rm v}$,
	\item maximizes $N_{\rm v}$ given the minimum $Z_{\rm v}$,
	\item maximizes $p_{n} = N_{\rm v} (1-Z_{\rm v})/2$, and 
	\item maximizes $p_{n} - p_{1} = -N_{\rm v} Z_{\rm v}$.
\end{itemize}

We may now prove that the optimal cycle achieves the largest change in the bias of an external qubit via the swap operation. Via \eqref{appbiaschange}, the difference in bias at the end of the swap is
\begin{equation}\label{swapbiaschange}
	Z_{\rm s}^\prime - Z_{\rm s} = N_{\rm v} (Z_{\rm v} - Z_{\rm s}).
\end{equation}

Labelling the norm and bias of the optimal $n$ level fridge as $\{N_{\rm v}^+,Z_{\rm v}^+\}$, and that of an arbitrary $n$ level cycle as $\{N_{\rm v}, Z_{\rm v}\}$,
\begin{align}
	Z_{\rm v} &\leq Z_{\rm v}^+ & N_{\rm v} Z_{\rm v} &\leq N_{\rm v}^+ Z_{\rm v}^+.
\end{align}

Thus for the swap using an arbitrary cycle,
\begin{align}
	Z_{\rm s}^\prime - Z_{\rm s} &< \frac{N_{\rm v}^+ Z_{\rm v}^+}{Z_{\rm v}} \left( Z_{\rm v} - Z_{\rm s} \right) = N_{\rm v}^+ Z_{\rm v}^+ \left( 1 - \frac{Z_{\rm s}}{Z_{\rm v}} \right), \nonumber\\
	&< N_{\rm v}^+ Z_{\rm v}^+ \left( 1 - \frac{Z_{\rm s}}{Z_{\rm v}^+} \right) = N_{\rm v}^+ \left( Z_{\rm v}^+ - Z_{\rm s} \right).
\end{align}

Thus the change in the bias is upper bounded by that achieved by the optimal fridge cycle. One may also prove the analogous result involving the optimal engine cycle,
\begin{equation}
	Z_{\rm s} - Z_{\rm s}^\prime = N_{\rm v} (Z_{\rm s} - Z_{\rm v}) < N_{\rm v}^- \left( Z_{\rm s} - Z_{\rm v}^- \right),
\end{equation}
where $\{N_{\rm v}^-,Z_{\rm v}^-\}$ are the norm and bias of the optimal engine cycle.

\bigskip
\subsection{Efficiency of single cycle machines}

Recall the normal definitions of efficiency for absorption thermal machines. For a fridge, this is defined as the ratio between the heat drawn from the object to be cooled to the heat drawn from the hot bath. For an engine, it is the ratio between the work done to the heat drawn from the hot bath.

In the case of the thermal cycle, the energy gap of the virtual qubit $E_{\rm v}$ represents both the heat drawn in the case of the fridge, and the work done in the case of the engine. Every time the virtual qubit exchanges $E_{\rm v}$ with an external system, it has to be reset by moving through the entire cycle. From Table \ref{transitionsizeintermediate}, we can calculate the heat currents to/from each bath, identifying $Q_+$ and $Q_-$ from the table with $Q_{\rm c}$ and $Q_{\rm h}$ respectively, in the case of the fridge, and the opposite for the engine.

One can thus re-express the virtual temperature of the thermal cycle (Table \ref{optimumbetav}) in terms of the heat currents,
\begin{align}
 \text{(fridge)   }   \quad	\beta_{v}^{(n)} E_{\rm v} &= \beta_{\rm c} \left( Q_{\rm h} + E_{\rm v} \right) - \beta_{\rm h} Q_{\rm h}, \\
\text{(engine)   }   \quad	\beta_{v}^{(n)} E_{\rm v} &= \beta_{\rm h} Q_{\rm h}- \beta_{\rm c} \left( Q_{\rm h} - E_{\rm v} \right).
\end{align}

Solving for the efficiency $\eta = E_{\rm v}/Q_{\rm h}$, one recovers the efficiencies of the thermal cycle,
\begin{align}
	\eta_{fridge}^{(n)} &= \frac{\beta_{\rm c} - \beta_{\rm h}}{\beta_{\rm v}^{(n)} - \beta_{\rm c}}, & \eta_{engine}^{(n)} &= \frac{\beta_{\rm c} - \beta_{\rm h}}{\beta_{\rm c} - \beta_{\rm v}^{(n)}}.
\end{align}

In both cases the efficiency falls off with increasing $\beta_{\rm v}$, and thus in the case of the optimal $n$ level cycle, one finds that with an increasing number of levels, as the magnitude of $\beta_{\rm v}^{(n)}$ increases linearly with $n$, so the efficiency $\eta$ falls off inversely with $n$.

\onecolumngrid
\newpage
\section{Switching between fridges and engines}\label{switching}
\twocolumngrid

When viewed in reverse, the amplification of the norm of a virtual qubit (Section \ref{sec:multi-cycle}) presents itself as a novel method to amplify the norm of any thermal cycle to one; simply connect its virtual qubit to a real qubit via a suitable interaction Hamiltonian, and use the real qubit instead to interact with the external system. The real qubit is now our ``virtual qubit".

To be more precise, consider that one has a single $n-$level cycle, whose virtual qubit, labelled by the states $\ket{1}_{cycle}$ and $\ket{n}_{cycle}$, has an energy gap of $E_{\rm v}$ and a virtual temperature of $\beta_{\rm v}$.

Couple this transition to a real qubit (labelled by $\ket{g}_{\rm v}$ and $\ket{e_{\rm v}}$) with the same energy gap $E_{\rm v}$ via a swap-like Hamiltonian, such as
\begin{equation}
	H_{int} = g \left( \ket{1}_{cycle}\!\bra{n} \otimes \ket{e}_{\rm v}\!\bra{g } + c.c. \right).
\end{equation}

\begin{figure}[h]
\includegraphics[width=\linewidth]{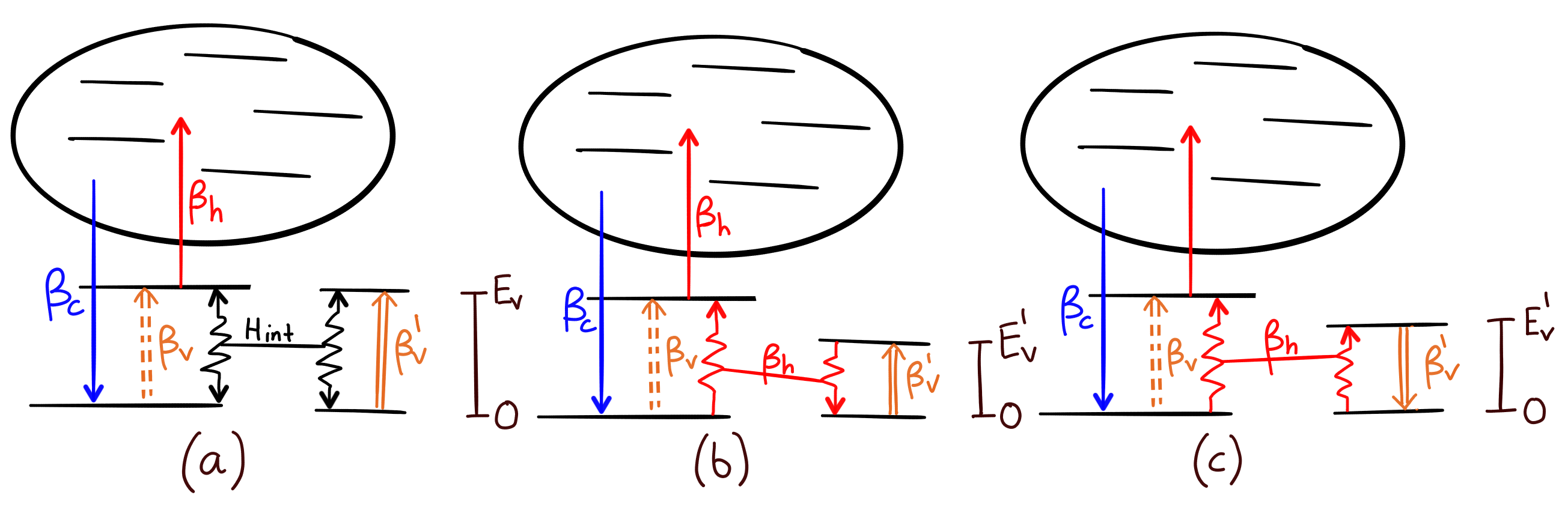}
\caption{Different methods of amplifying the virtual qubit of an arbitrary cycle. (a) Amplification that maintains the energy and bias of the virtual qubit. (b) Amplification that modifies (possibly amplifies) the bias of the virtual qubit. (c) Amplication that flips the bias of the virtual qubit.\label{realqubitcouplings}}
\end{figure}

This arrangement is depicted in Fig. \ref{realqubitcouplings}(a). In the steady state, the populations of the levels must satisfy
\begin{equation}
	p(\ket{1}_{cycle} \otimes \ket{e}_{\rm v}) = p(\ket{n}_{cycle} \otimes \ket{g}_{\rm v}).
\end{equation}

But since $p_{n}/p_{1} = e^{-\beta_{\rm v} E_{\rm v}}$ via the thermal cycle, it follows that the real qubit levels exhibit the same population ratio, i.e.
\begin{equation}
	\frac{p_{e_{\rm v}}}{p_{g_{\rm v}}} = e^{-\beta_{\rm v} E_{\rm v}}.
\end{equation}

This completes the virtual qubit amplification procedure, since $N_{\rm v}=1$ for the real qubit.

In fact one can do even more, if the states $\ket{1}_{cycle} \otimes \ket{e}_{\rm v}$ and $\ket{n}_{cycle} \otimes \ket{g}_{\rm v}$ are coupled via a thermal bath rather than an energy conserving interaction. In this case the two states need not be degenerate. If the energy gap of the real qubit is labelled as $E_{\rm v}^\prime$, and the two states above are coupled to $\beta_{bath}$, as in Fig. \ref{realqubitcouplings}(b), then in the steady state, the populations satisfy
\begin{align}
	\frac{p_{1} p_{e_{\rm v}}}{p_{n} p_{g_{\rm v}}} &= e^{-\beta_{bath} (E_{\rm v}^\prime - E_{\rm v})}.
\end{align}

Once again the virtual temperature of the virtual qubit of the cycle, $p_{n}/p_{1} = e^{-\beta_{\rm v} E_{\rm v}}$, the virtual temperature $\beta_{\rm v}^\prime$ of the real qubit may be determined,
\begin{equation}
	\beta_{\rm v}^\prime E_{\rm v}^\prime = \beta_{\rm v} E_{\rm v} + \beta_{bath} (E_{\rm v}^\prime - E_{\rm v}).
\end{equation}

Finally, consider that rather than couple the states $\ket{1}_{cycle} \otimes \ket{e}_{\rm v}$ and $\ket{n}_{cycle} \otimes \ket{g}_{\rm v}$, one couples instead $\ket{1}_{cycle} \otimes \ket{g}_{\rm v}$ and $\ket{n}_{cycle} \otimes \ket{e}_{\rm v}$ to a thermal bath, see Fig. \ref{realqubitcouplings}(c). Similarly to the above, one may determine that the real qubit has the virtual temperature
\begin{equation}
	\beta_{\rm v}^\prime E_{\rm v}^\prime = -\beta_{\rm v} E_{\rm v} + \beta_{bath} (E_{\rm v} + E_{\rm v}^\prime).
\end{equation}
However, in this case the contribution of the original virtual temperature is multiplied by $-1$, effectively switching the machine from a fridge to an engine or vice versa! Thus given a $n-$level fridge cycle, one may switch to an engine and vice-versa, by using the appropriate thermal coupling between the cycle and the real qubit.

\onecolumngrid

\twocolumngrid

\onecolumngrid
\newpage

\section{Concatenated qutrit machines}\label{qutrits}

\begin{figure}[h]
\includegraphics[width=0.9\linewidth]{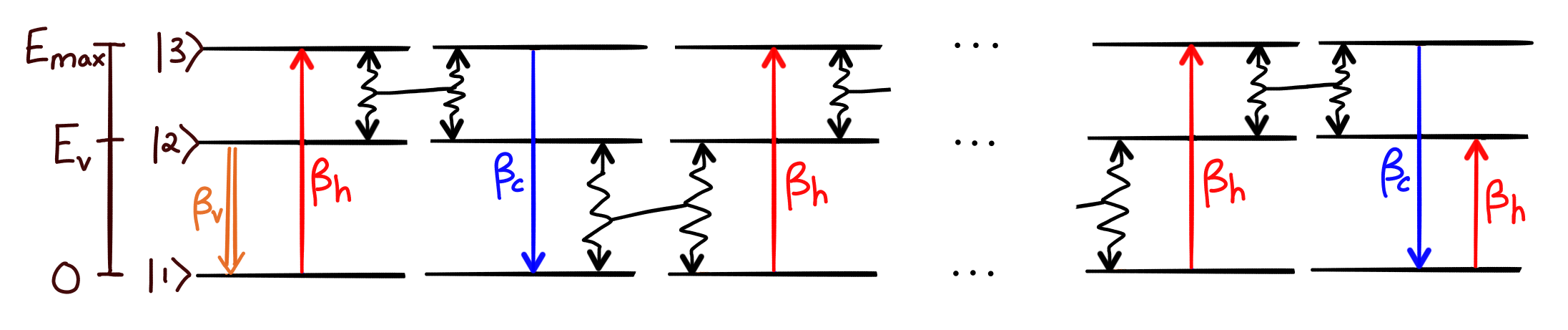}
\caption{Engine formed out of the concatenation of many qutrit machines.\label{nqutritappendix}}
\end{figure}

\twocolumngrid
In this section we consider the concatenation of qutrit machines, and determine the bias and norm of the virtual qubit in its steady state of operation.

To arrive at the steady state, it is simpler to begin from the end of the concatenation, and derive the state inductively. To begin with, consider the final (rightmost) qutrit in Fig. \ref{nqutritappendix}, ignoring it's interaction with the penultimate qutrit. It is equivalent to a single qutrit fridge, and it's populations are completely determined by the two thermal couplings.

One now introduces a swap-like interaction between the uncoupled transition of the final qutrit and the corresponding transition of the penultimate qutrit,
\begin{equation}
	H_{int} = g ( \ket{12}_{n(n-1)}\bra{21} + c.c. ).
\end{equation}

This interaction induces the transition of the penultimate qutrit $\Gamma_{12}^{(n-1)}$ to have the same Gibbs ratio as that of $\Gamma_{12}^{(n)}$.

If one also couples $\Gamma_{13}^{(n-1)}$ to $\beta_{\rm h}$, that fixes a second Gibbs ratio on the penultimate qutrit, leading to the populations of the penultimate qutrit being completely determined. The state is still diagonal, and a product state, as the thermal couplings only fix the Gibbs ratio on single qutrits, while the interaction matches the Gibbs ratio of a transition whose ratio is already fixed, to one that is not yet determined.

Note that the same state of the penultimate qutrit would have been found if one had simply assumed that in place of the final qutrit, there was instead a thermal bath at the virtual temperature of $\Gamma_{12}^{(n)}$.

One may repeat this process inductively to determine the state of the first qutrit in the sequence, and in turn the virtual temperature of the transition $\Gamma_{01}^{(1)}$, finding as in the main text \eqref{nqutrit}
\begin{equation}\label{nqutritapp}
	\beta_{\rm v}^{(k)} =  
		\begin{cases}
		\beta_{\rm c} + (\beta_{\rm c} - \beta_{\rm h})\frac{k}{2}\frac{E_{\rm max}}{E_{\rm v}} & \text{ if $k$ is even,} \\
		\beta_{\rm c} + (\beta_{\rm c} - \beta_{\rm h})\left( \frac{k+1}{2}\frac{E_{\rm max}}{E_{\rm v}} - 1 \right) & \text{ if $k$ is odd.}
		\end{cases}
\end{equation}

The virtual temperatures for the engine are the same as above with $\beta_{\rm c}$ and $\beta_{\rm h}$ switched. Note that the virtual temperature of $k$ concatenated qutrit is identical to that of the optimal $k+2$ level thermal cycle, Table \ref{optimumbetav}.

We are also interested in calculating the norm $N_{\rm v}$  of the virtual qubit. An interesting freedom in the case of the qutrit machine is the choice of whether to have the virtual qubit as the transition between the lower two levels $\Gamma_{12}$ or $\Gamma_{23}$ of the first qutrit (modifying the energies accordingly so that the energy gap is always $E_{\rm v}$). We are especially interested in the behaviour of the norm as the number of concatenated qutrits becomes large (and $\beta_{\rm v}$ approaches $\pm \infty$.)

While this choice has no bearing on the bias of the virtual qubit, it does affects its norm. On may calculate for the case of the fridge, the norm of the virtual qubit is
\begin{align}
	N^{(23)}_{\rm v} &= \frac{1 + e^{-\beta_{\rm v} W}}{1 + e^{-\beta_{\rm v} W} + e^{-\beta_{\rm v} W} e^{+\beta_{\rm c} E_{\rm max}}} \\
	\lim_{\beta_{\rm v} \rightarrow +\infty} N^{(23)}_{\rm v} &= 1,
\end{align}
in the case that the virtual qubit is $\Gamma_{23}$, and
\begin{align}
	N^{(12)}_{\rm v} &= \frac{1 + e^{-\beta_{\rm v} W}}{1 + e^{-\beta_{\rm v} W} + e^{-\beta_{\rm c} E_{\rm max}}}, \\
	\lim_{\beta_{\rm v} \rightarrow +\infty} N^{(12)}_{\rm v} &= \frac{1}{1 + e^{-\beta_{\rm c} E_{\rm max}}}
\end{align}
in the case the virtual qubit is $\Gamma_{12}$. Clearly it is advantageous to place the virtual qubit on the upper two levels.

This is the opposite for the case of the engine. We find that the corresponding norms for the case of lower and upper virtual qubits is respectively
\begin{align}
	N^{(23)}_{\rm v} &= \frac{1 + e^{+\beta_{\rm v} W}}{1 + e^{+\beta_{\rm v} W} + e^{+\beta_{\rm h} E_{\rm max}}}, \\
	\lim_{\beta_{\rm v} \rightarrow -\infty} N^{(23)}_{\rm v} &= \frac{1}{1 + e^{-\beta_{\rm h} E_{\rm max}}}. \\
	N^{(12)}_{\rm v} &= \frac{1 + e^{+\beta_{\rm v} W}}{1 + e^{+\beta_{\rm v} W} + e^{+\beta_{\rm v} W} e^{\beta_{\rm h} E_{\rm max}}}, \\
	\lim_{\beta_{\rm v} \rightarrow -\infty} N^{(12)}_{\rm v} &= 1.
\end{align}

This motivates the choice of $\Gamma_{23}$ as the virtual qubit for the fridge, and $\Gamma_{12}$ as the virtual qubit for the engine. Also note that via this choice, in the limit $n\rightarrow \infty$, both the fridge and the engine qutrits approach the same state, i.e. a qutrit with all of its population in the middle energy level.


\begin{thebibliography}{99}
\bibitem{footnote1} More generally, one could also consider the case of virtual qubits with coherences.
\bibitem{footnote3} If the cycle covers only a subspace of all the machine levels, then the populations are determined with respect to the total population of the subspace.
\end{thebibliography}
\end{document}